\documentclass[useAMS,usenatbib]{mn2e}
\input psfig.sty

\title[Companion Stars of Type Ia supernovae]
{Companion Stars of Type Ia supernovae with different
metallicities}
\author[Meng and Yang]
       {X. Meng$^{\rm 1}$ \thanks{E-mail:
conson859@msn.com} and W. Yang$^{\rm 1}$ \\
        $^1$ Department of Physics and Chemistry, Henan Polytechnic
University, Jiaozuo, 454000, China}

\begin{document}

\date{Accepted. Received}

\pagerange{\pageref{firstpage}--\pageref{lastpage}} \pubyear{2007}

\maketitle

\label{firstpage}

\begin{abstract}

Single degenerate model is the widely accepted progenitor model of
Type Ia supernovae (SNe Ia), where a CO WD accretes hydrogen-rich
material from its companion to increase its mass. The companion
may be a main-sequence star or a subgiant star (WD + MS). When the
CO WD approaches the Chandrasekhar mass limit, it explodes as a SN
Ia and a part of supernova ejecta collides into the companion
envelope. After the impact of the ejecta, the companion survives
and may show some special properties. A good way to verify the
single degenerate model is to study the interaction between
supernova ejecta and its companion, and/or search companion in the
remnant of a SN Ia.

\citet{MENGCH09} comprehensively and systemically studied the WD +
MS system by detailed binary evolution calculations. Following
their studies, we have carried out a series of binary population
synthesis studies about the properties of the companions of SNe Ia
for different metallicities $Z$. We present the distributions of
the masses, $M_{\rm 2}^{\rm SN}$, the radii, $R_{\rm 2}^{\rm SN}$,
of companions, and the periods, $P_{\rm SN}$, and the ratios of
separations to radii, $A/R_{\rm 2}^{\rm SN}$, of WD + MS systems
for various $Z$ at the moment of supernova explosion. These
parameters can be applied to constrain the numerical simulation of
the interaction between the ejecta of a supernova and its
companion. We also show the distributions of some integral
properties of companions, i.e. the mass, the space velocity and
the surface gravity, for various $Z$ after the interaction. The
distributions may help to search companion in a supernova remnant.
All the parameters above significantly change with $Z$.

Incorporating the simulation results of interaction between
supernovae ejecta and companions in \citet{MAR00} and
\citet{KAS04} into our binary population synthesis study, we found
that more than 75\% of all supernovae have enough polarization
signal which can be detected by spectropolarimetric observations.
We also found that 13 to 14 per cent SNe Ia belong to the
supernovae like 1991T, which is consistent with observations
within errors. This may indicate that SNe 1991T-like have not any
special properties in physics except for the viewing angle of an
observer.
\end{abstract}


\section{Introduction}\label{sect:1}
Type Ia supernovae (SNe Ia) play an important role in
astrophysics, especially in cosmology. They appear to be good
cosmological distance indicators and are successfully applied to
determine cosmological parameters (e.g. $\Omega$ and $\Lambda$;
\citealt{RIE98}; \citealt{PER99}). There is a linear relation
between the absolute magnitude of SNe Ia and the magnitude
difference from maximum to 15 days after B maximum light. The
relation is known as Phillips relation (\citealt{PHI93}) and
adopted when SNe Ia were taken as the distance indicators. In this
case, Phillips relation is assumed to be valid at high redshift,
although it was obtained from a low-redshift sample. This
assumption is precarious since the exact nature about SNe Ia is
still unclear (see the reviews by \citealt{HN00};
\citealt{LEI00}). If the properties of SNe Ia evolve with
redshift, the results for cosmology might be different. Since
metallicity may represents redshift to some extent, it is a good
method to study the properties of SN Ia at various redshift by
finding the correlation between the properties and metallicity.

It is widely believed that SN Ia is from the thermonuclear runaway
of a carbon-oxygen white dwarf (CO WD) in a binary system. The CO
WD accretes material from its companion to increase its mass. When
its mass reaches its maximum stable mass, it explodes as a
thermonuclear runaway and almost half of the WD mass is converted
into radioactive nickel-56 (\citealt{BRA04}). The mass of
nickel-56 determines the maximum luminosity of SN Ia. The higher
the mass of nickel-56 is, the higher the maximum luminosity is
(\citealt{ARN82}). Some numerical and synthetical results have shown
that metallcity may affect the final amount of nickel-56, and thus
the maximum luminosity (\citealt{TIM03}; \citealt{TRA05};
\citealt{POD06}). There is also much evidence about the
correlation between the properties of SNe Ia and metallicity in
observations (we list here some of the papers on this study.
\citealt{BB93}; \citealt{HAM96}; \citealt{WAN97}; \citealt{CAP97};
\citealt{SHA02}).

Among all the suggested progenitor models of SNe Ia, the
single-degenerate Chandrasekhar model (\citealt{WI73};
\citealt{NTY84}) is the most widely accepted progenitor model at
present. In the model, a CO WD accretes hydrogen-rich material
from its companions until its mass reaches a mass $\sim 1.378
M_{\odot}$ (close to Chandrasekhar mass, \citealt{NTY84}), and
then explodes as a SN Ia. The companion may be a main sequence
star or a subgiant star (WD+MS) or a red-giant star (WD+RG)
(\citealt{YUN95}; \citealt{LI97}; \citealt{HAC99a, HAC99b};
\citealt{NOM99, NOM03}; \citealt{LAN00}; \citealt{HAN04}; \citealt{CHENWC07, CHENWC09};
\citealt{HAN08} \citealt{MENGCH09}; \citealt{LGL09}; \citealt{WANGB09a, WANGB09b}). In this
paper, we only focus on the WD + MS channel since it is
the most widely accepted channel for SNe Ia
(\citealt{HAN04}; \citealt{MENGCH09}). Much observational evidence
shows the importance of the channel. For example, some WD + MS
systems are suggested as the progenitor of SNe Ia
(\citealt{PAR07}). \citet{HK03a, HK03b} suggested that supersoft
X-ray sources (SSSs) may be good candidates for the progenitors of
SNe Ia, and some of SSSs belong to WD+MS system. The discovery of
the potential companion of Tycho's supernova seemed to verify the
reliability of the WD + MS model (named Tycho G by
\citealt{RUI04}). Recently, \citet{HERNANDEZ09} stressed further
the companion nature of Tycho G by analysing the chemical
abundances of Tycho G. However, \citet{FUH05} argued that the
first discovered ``companion" might be a Milky Way thick-disk star
which is coincidentally passing the vicinity of the remnant of
Tycho's supernova. \citet{IHA07} also argued that Tycho G may not
be the companion of Tycho's supernova since the star did not show
any special properties in its spectrum, which should be
contaminated by supernova ejecta and show some special characters
(\citealt{MAR00}; \citealt{BRA04}). So, more evidence to confirmed
the companion nature of Tycho G is needed.

The knowledge about the companions of SNe Ia after explosions is
still unclear. Generally, the supernova ejecta in the single
degenerate model collides into the envelope of its companion and
strips some hydrogen-rich material from the surface of the
companion. After the collision, the companion gains a kick
velocity, which is much smaller than orbital velocity, and leaves
explosion center at a velocity similar to its orbital velocity
(\citealt{CHE74}; \citealt{WHE75}; \citealt{FRY81};
\citealt{TAA84}; \citealt{CHU86}; \citealt{LIV92};
\citealt{LAN00}). \citet{MAR00} ran several high-resolution
two-dimensional numerical simulations of the collision between the
ejecta and the companion, where the companion is a MS star, a
subgiant (SG) star or a red giant (RG) star. They found that about
0.15 $M_{\rm \odot}$ - 0.17 $M_{\rm \odot}$ of hydrogen-rich
materials are stripped from the surface of a MS or a SG companion
and n a sense of the collision, there is no difference
whatever the companion is a MS star or a SG star. The amount of
the stripped material from red-giant companions is even more than
$0.5 M_{\odot}$, i.e. more than 96\% of their envelopes is lost.
\citet{MENGCH07} used a simple analytic method but a more physical
companion model to simulate the interaction and found that the
minimum value of the stripped material from a MS companion is
diminished from 0.15 $M_{\rm \odot}$ to 0.035 $M_{\rm \odot}$.
They suggested that the structure of the companion at the
explosion moment may be an important factor to determine
the mass of the stripped material. The structure of the companion
in \citet{MENGCH07} is obviously different from that of a solar
model used in \citet{MAR00} since the companion still do not reach
thermal equilibrium because of mass transfer. The
reduction of the stripped material in \citet{MENGCH07} compared
with that in \citet{MAR00} primarily results from the
pre-explosion mass loss. The MS model in \citet{MENGCH07} is from
a more massive star experiencing a mass-loss phase before
supernova explosion, which leads to a more compact companion star
whose material is more difficult to strip than it is in a solar
model as used in \citet{MAR00}. Since \citet{MENGCH07} did not
consider the thermal energy imparted by the ejecta into the
companion envelope, which likely heats and vaporizes a part of the
envelope and thereby increases the amount of stripped material,
0.035 $M_{\rm \odot}$ should be a conservative lower limit. If
enough hydrogen-rich materials are stripped, they should reveal
themselves by narrow H$_{\rm \alpha}$ emission or absorption line
in later-time spectra of SNe Ia (\citealt{CHU86};
\citealt{FIL97}). However, the H$_{\rm \alpha}$ line was not
detected and the amount of the stripped material was constrained
to be less than 0.02 $M_{\odot}$ by observations (\citealt{MAT05};
\citealt{LEO07}), which is much smaller than 0.15 $M_{\rm \odot}$
- 0.17 $M_{\rm \odot}$ and is even smaller than the lower limit
obtained by \citet{MENGCH07}. In addition, the simulation in
\citet{MAR00} and \citet{MENGCH07} showed that the luminosity of
supernova companion after collision by supernova ejecta should
dramatically rise to a level much higher than that of Tycho G by
about three orders of magnitude.

Owing to the discussions above, a detailed numerical simulation on
the interaction between supernova ejecta and its companion should
be very important, while the subject is closely related with the
properties of the secondaries before SNe Ia explosion. Recently,
\citet{MENGCH09} (hereinafter Paper I) calculated a dense grids of
binary systems by detailed binary evolution for different
metallicities and gave the initial parameters of the systems
leading to SNe Ia in an orbital period-secondary mass ($\log
P_{\rm i}, M_{\rm 2}^{\rm i}$) plane. The aim of this paper is to,
using the results in Paper I, show the properties of secondaries
before and after the SNe Ia explosion, which can provide help to
do detailed numerical simulations of the collision between
supernova ejecta and its companion, and/or search the companion in
the explosion remnant of a SN Ia.

The paper is organized as follow. We simply show our binary
population synthesis (BPS) method in section \ref{sect:2} and the
BPS results in section \ref{sect:3}. In section \ref{sect:4}, we
briefly discuss our results and summarize the main conclusions in
section \ref{sect:5}.
\begin{figure}
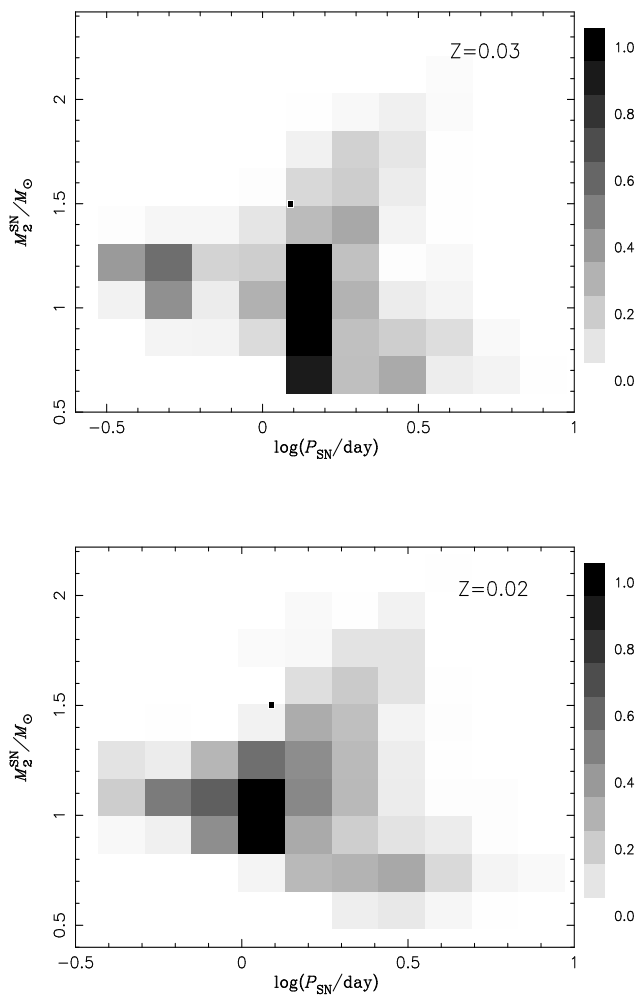

 \centering
\centerline{\psfig{figure=permms0310.ps,width=0.3\textheight,height=0.40\textwidth,bbllx=60pt,bblly=100pt,bburx=585pt,bbury=676pt,angle=270}}
\centerline{\psfig{figure=permms0210.ps,width=0.3\textheight,height=0.40\textwidth,bbllx=60pt,bblly=100pt,bburx=585pt,bbury=676pt,angle=270}}
 \suppressfloats[t]
 \caption{The distribution of the final masses and the orbital
periods at the moment of explosions. Common envelope ejection
efficiency $\alpha_{\rm CE}=1.0$. The position of a recurrent
nova, U Sco, is indicated by the filled square (\citealt{SR95};
\citealt{HK00a, HK00b}). Top: $Z=0.03$; Bottom: $Z=0.02$.}
 \label{permms0302}
\end{figure}

\begin{figure}
 \centering
\centerline{\psfig{figure=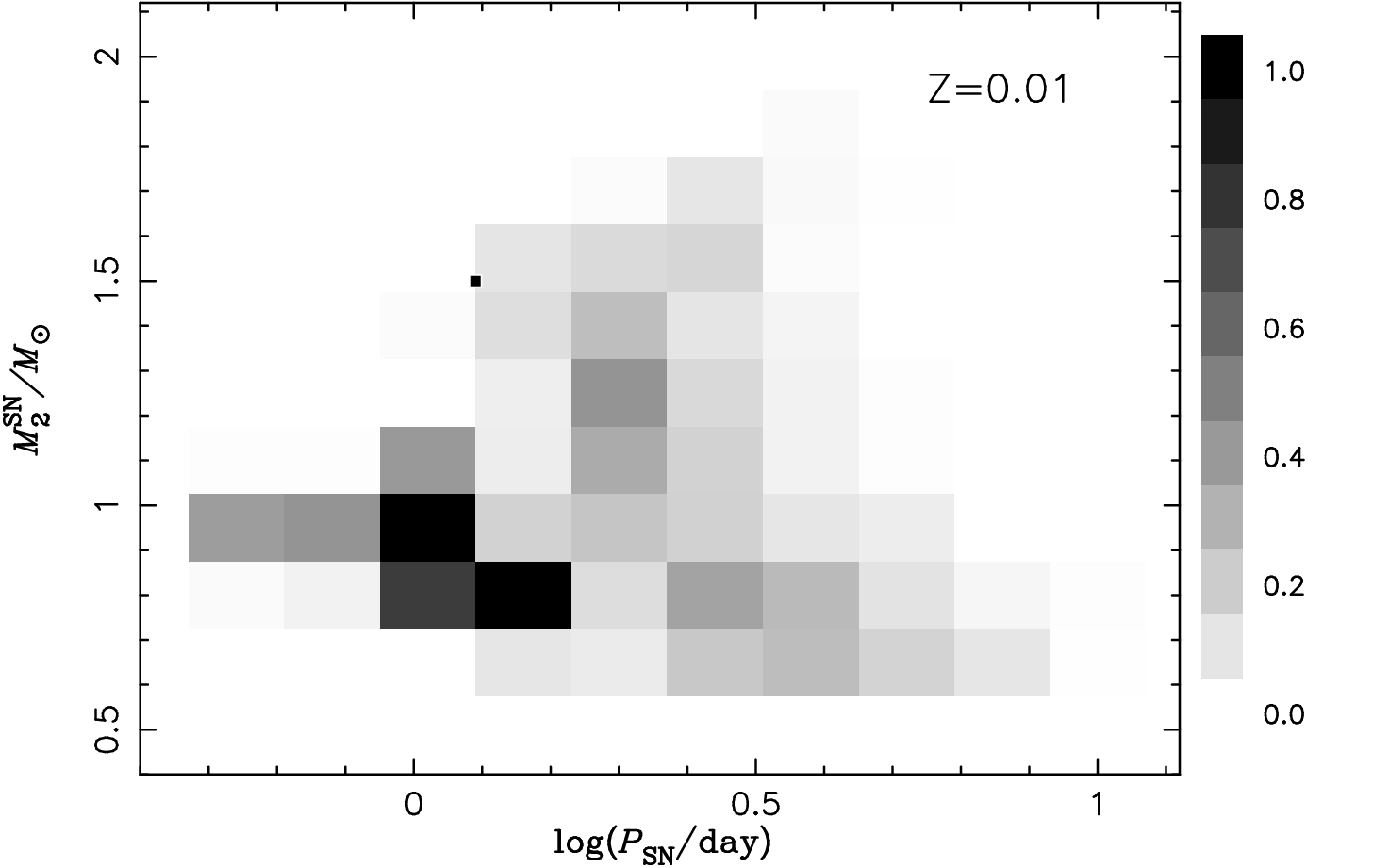,width=0.3\textheight,height=0.40\textwidth,bbllx=60pt,bblly=100pt,bburx=585pt,bbury=676pt,angle=270}}
\centerline{\psfig{figure=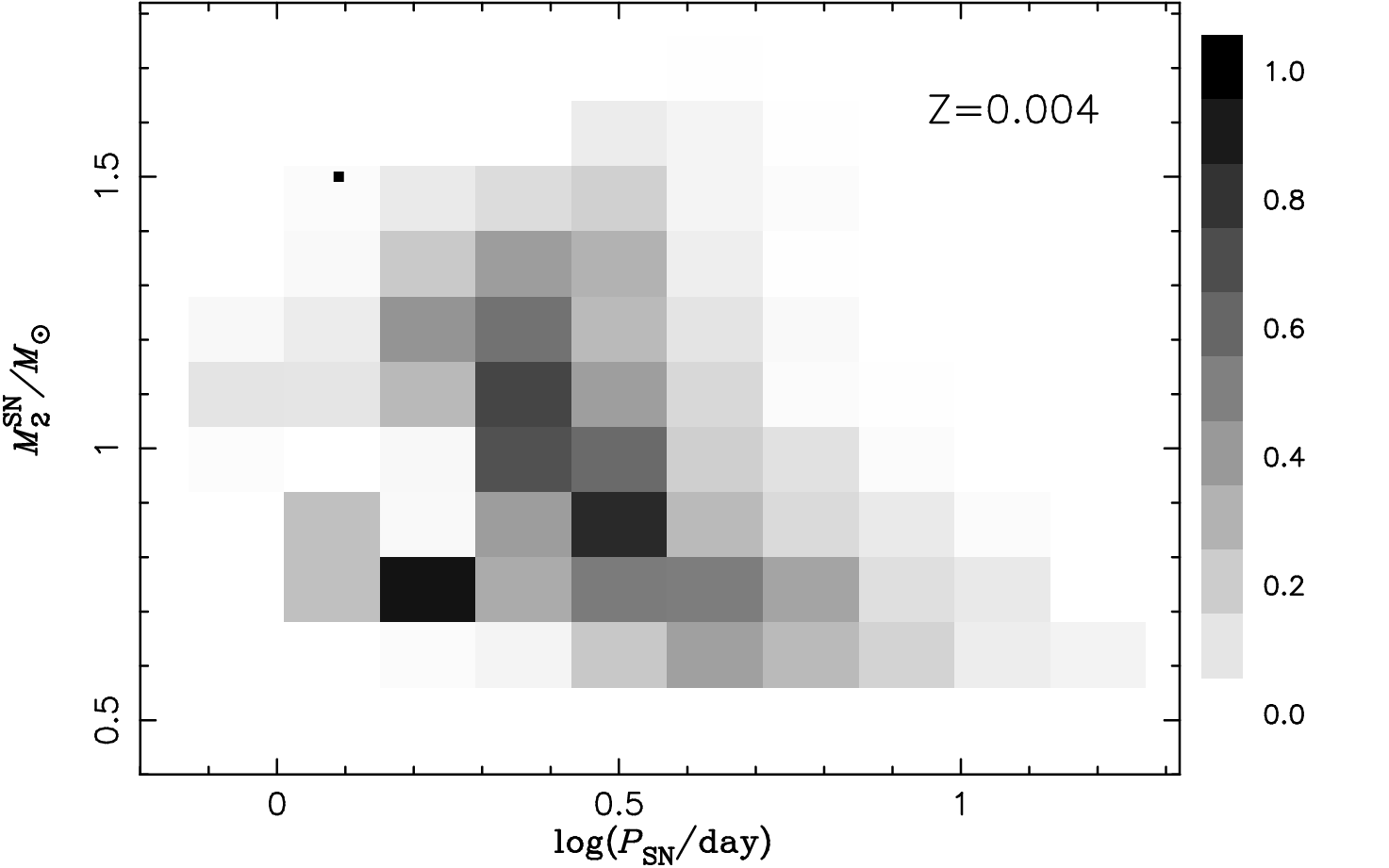,width=0.3\textheight,height=0.40\textwidth,bbllx=60pt,bblly=100pt,bburx=585pt,bbury=676pt,angle=270}}
 \suppressfloats[t]
 \caption{Similar to Fig \ref{permms0302} but for $Z=0.01$ (Top) and $Z=0.004$ (Bottom).}
 \label{permms01004}
\end{figure}

\begin{figure}
 \centering
\centerline{\psfig{figure=permms00110.ps,width=0.3\textheight,height=0.40\textwidth,bbllx=60pt,bblly=100pt,bburx=585pt,bbury=676pt,angle=270}}
\centerline{\psfig{figure=permms000310.ps,width=0.3\textheight,height=0.40\textwidth,bbllx=60pt,bblly=100pt,bburx=585pt,bbury=676pt,angle=270}}
 \suppressfloats[t]
 \caption{Similar to Fig \ref{permms0302} but for $Z=0.001$ (Top) and $Z=0.0003$ (Bottom).}
 \label{permms0010003}
\end{figure}

\begin{figure}
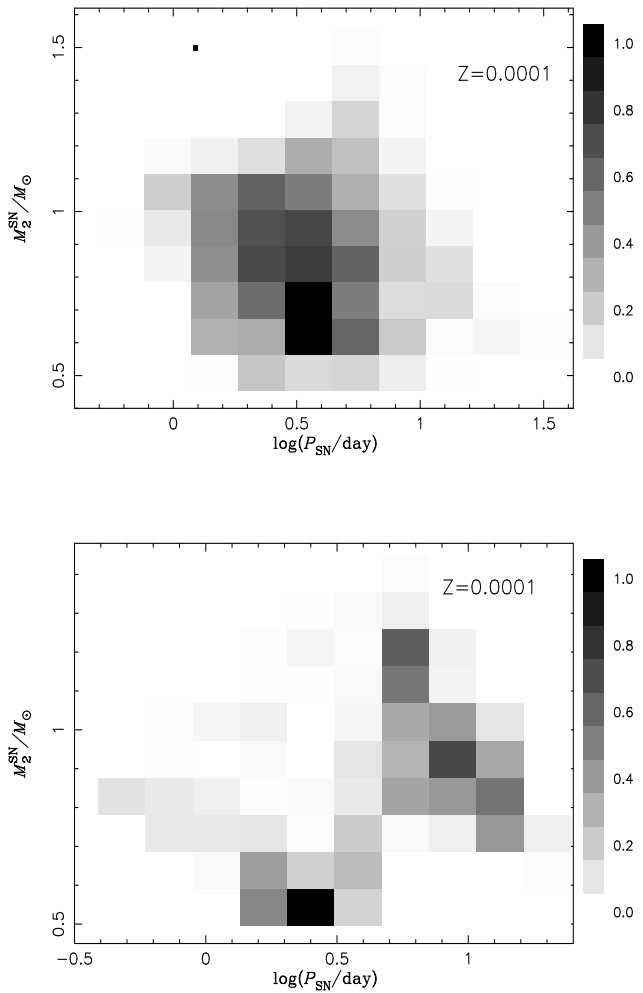

 \centering
\centerline{\psfig{figure=permms000110.ps,width=0.3\textheight,height=0.40\textwidth,bbllx=60pt,bblly=100pt,bburx=585pt,bbury=676pt,angle=270}}
\centerline{\psfig{figure=permms000130.ps,width=0.3\textheight,height=0.40\textwidth,bbllx=60pt,bblly=100pt,bburx=585pt,bbury=676pt,angle=270}}
 \suppressfloats[t]
 \caption{Similar to Fig \ref{permms0302} but for $Z=0.0001$. Top: $\alpha_{\rm CE}=1.0$; Bottom: $\alpha_{\rm CE}=3.0$.}
 \label{permms00010001}
\end{figure}

\section[]{BINARY POPULATION SYNTHESIS}\label{sect:2}

\subsection{The parameters of WD+MS systems leading to SNe Ia}\label{subs:2.1}
Incorporating the prescription  in \citet{HAC99a} on the accretion
of the hydrogen-rich material from its companion onto a WD into
the Eggleton's stellar evolution code (\citealt{EGG71, EGG72,
EGG73}), \citet{MENGCH09} calculated more than 25,000 binary
evolutions of WD+MS channel. In the channel, a companion fills its
Roche lobe in MS stage or in Hertzprung gap (HG) and transfers
some of its mass onto the surface of the WD. As a consequence, the
mass of the WD increases gradually. If the mass of the WD reaches
$\sim 1.378 M_{\odot}$ (\citealt{NTY84}), they assumed that the WD
explodes as a SN Ia. \citet{MENGCH09} provided the initial
parameter spaces of the WD + MS systems leading to SNe Ia in an
orbital period - secondary mass ($\log P_{\rm i}, M_{\rm 2}^{\rm
i}$) plane. Their calculations are comprehensive and systematical,
and various properties of the companion stars with different
metallicities were obtained but not sorted for publishing. In this
paper, we extract the properties from the data files of the
calculations and incorporate them into the rapid binary evolution
code developed by \citet{HUR00, HUR02} to obtain the various
properties of the companions at the moment of explosion.

\subsection{Common envelope}\label{subs:2.2}
Common envelope (CE) is very important for the formation of WD +
MS systems. We firstly introduce the treatment for CE in this
paper. Hereinafter, we use {\sl primordial} to represent the stage
before the formation of WD+MS systems and {\sl initial} for WD+MS
systems. During binary evolution, the primordial mass ratio
(primary to secondary) is crucial for the first mass transfer. If
it is larger than a critical mass ratio, $q_{\rm c}$, the first
mass transfer is dynamically unstable and a CE forms
(\citealt{PAC76}). The ratio $q_{\rm c}$ varies with the
evolutionary state of the primordial primary at the onset of RLOF
(\citealt{HW87}; \citealt{WEBBINK88}; \citealt{HAN02};
\citealt{POD02}; \citealt{CHE08}). In this study, we adopt $q_{\rm
c}$ = 4.0 when the primary is on MS or HG. This value is supported
by detailed binary evolution studies (\citealt{HAN00};
\citealt{CHE02, CHE03}). If the primordial primary is on FGB or
AGB, we use

\begin{equation}
q_{\rm c}=[1.67-x+2(\frac{M_{\rm c1}^{\rm P}}{M_{\rm 1}^{\rm
P}})^{\rm 5}]/2.13,  \label{eq:qc}
  \end{equation}
where $M_{\rm c1}^{\rm P}$ is the core mass of primordial primary,
and $x={\rm d}\ln R_{\rm 1}^{\rm P}/{\rm d}\ln M_{\rm 1}^{\rm p}$
is the mass-radius exponent of primordial primary and varies with
composition. If the mass donors (primaries) are naked helium
giants, $q_{\rm c}$ = 0.748 based on equation (\ref{eq:qc}) (see
\citealt{HUR02} for details).

Embedded in the CE are the dense core of the primordial primary
and the primordial secondary. Due to frictional drag with the
envelope, the orbit of the embedded binary decays and a large part
of the orbital energy released in the spiral-in process is
injected into the envelope (\citealt{LS88}). Here, we assume that
the CE is ejected if
\begin{equation}
\alpha_{\rm CE}\Delta E_{\rm orb}\geq |E_{\rm bind}|,
  \end{equation}
where $\Delta E_{\rm orb}$ is the orbital energy released, $E_{\rm
bind}$ is the binding energy of common envelope, and $\alpha_{\rm
CE}$ is CE ejection efficiency, i.e. the fraction of the released
orbital energy used to eject the CE. Since the thermal energy in
the envelope is not incorporated into the binding energy,
$\alpha_{\rm CE}$ may be greater than 1 (see \citealt{HAN95} for
details about the thermal energy). In this paper, we set
$\alpha_{\rm CE}$ to 1.0 or 3.0.

\subsection{Evolution channels}\label{subs:2.3}

There are three channels to produce WD + MS systems according to
the situation of the primary in a primordial system at the onset
of the first RLOF.

Case 1 (He star channel): the primordial primary is in HG or on
RGB at the onset of the first RLOF (i.e. case B evolution defined
by \citealt{KW67}). In this case, a CE is formed because of a
large mass ratio or a convective envelope of the mass donor. After
the CE ejection (if it occurs), the mass donor becomes a helium
star and continues to evolve. The helium star likely fills its
Roche lobe again after the central helium is exhausted. Since the
mass donor is much less massive than before, this RLOF is
dynamically stable, resulting in a close CO WD+MS system (see
\citealt{NOM99, NOM03} for details).

Case 2 (EAGB channel): the primordial primary is in early
asymptotic giant branch stage (EAGB) (i.e. helium is exhausted in
the core, while thermal pulses have not yet started). A CE is
formed because of dynamically unstable mass transfer. After the CE
is ejected, the orbit decays and the primordial primary becomes a
helium red giant (HeRG). The HeRG may fill its Roche lobe and
start the second RLOF. Similar to the He star channel, this RLOF
is stable and produces WD + MS systems after RLOF.

Case 3 (TPAGB channel): the primordial primary fills its Roche
lobe at the thermal pulsing AGB (TPAGB) stage. Similar to the
above two channels, a CE is formed during the RLOF. A CO WD + MS
binary is produced after CE ejection.

The WD + MS systems continue to evolve and the secondaries may
also fill their Roche lobes at a stage and Roche lobe overflow
(RLOF) starts. We assume that if the initial orbital period,
$P_{\rm orb}^{\rm i}$, and the initial secondary mass, $M_{\rm
2}^{\rm i}$, of a WD + MS system locate in the appropriate regions
in the ($\log P^{\rm i}, M_{\rm 2}^{\rm i}$) plane for SNe Ia at
the onset of RLOF (see Paper I), a SN Ia is then produced. The
properties of the binary system at the moment of SN explosion are
obtained by interpolation in the three-dimensional grid ($M_{\rm
WD}^{\rm i}, M_{\rm 2}^{\rm i}, \log P^{\rm i}$) of the more than
25,000 close WD binary system calculated in Paper I.

\subsection{Basic parameters in Monte Carlo simulation}\label{subs:2.4}
To investigate the statistical properties of companions before and
after explosion, we carry out a series of binary population
synthesis (BPS) studies for various $Z$ by Hurley's rapid binary
evolution code (\citealt{HUR00, HUR02}) incorporating the results
in Paper I. Since the code is only valid for $Z\leq0.03$, only
seven metallicities, i.e. $Z=0.03$, $0.02$, $0.01$, $0.004$,
$0.001$, $0.0003$ and $0.0001$, are examined here. In each BPS
study, $10^{\rm 7}$ binaries are generated by Monte Carlo
simulation and a circular orbit is assumed for all binaries. The
basic parameters for the simulations are as follows.

(i) The initial mass function (IFM) of \citet{MS79} is adopted.
The primordial primary is generated according to the formula of
\citet{EGG89}
\begin{equation}
M_{\rm 1}^{\rm p}=\frac{0.19X}{(1-X)^{\rm 0.75}+0.032(1-X)^{\rm
0.25}},
  \end{equation}
where $X$ is a random number in the range [0,1] and $M_{\rm
1}^{\rm p}$ is the mass of the primordial primary, which ranges
from 0.1 $M_{\rm \odot}$ to 100 $M_{\rm \odot}$.

(ii) The mass ratio of the primordial components, $q$, is a very
important parameter for binary evolution while its distribution is
quite controversial. For simplicity, we take a uniform mass-ratio
distribution (\citealt{MAZ92}; \citealt{GM94}):
\begin{equation}
n(q)=1, \hspace{2.cm} 0<q\leq1,
\end{equation}
where $q=M_{\rm 2}^{\rm p}/M_{\rm 1}^{\rm p}$.

(iii) We assume that all stars are members of binary systems and
that the distribution of separations is constant in $\log a$ for
wide binaries and falls off smoothly at close separation:
\begin{equation}
an(a)=\left\{
 \begin{array}{lc}
 \alpha_{\rm sep}(a/a_{\rm 0})^{\rm m} & a\leq a_{\rm 0};\\
\alpha_{\rm sep}, & a_{\rm 0}<a<a_{\rm 1},\\
\end{array}\right.
\end{equation}
where $\alpha_{\rm sep}\approx0.070$, $a_{\rm 0}=10R_{\odot}$,
$a_{\rm 1}=5.75\times 10^{\rm 6}R_{\odot}=0.13{\rm pc}$ and
$m\approx1.2$. This distribution implies that the numbers of wide
binary system per logarithmic interval are equal, and that
approximately 50\% of the stellar systems are binary systems with
orbital periods less than 100 yr (\citealt{HAN95}).

(iv)We simply assume a constant star formation rate (SFR) over last 15 Gyr.


\section{BINARY POPULATION SYNTHESIS RESULTS}\label{sect:3}
\subsection{\textbf{the companion masses and the orbital periods of WD + MS systems at the moment of explosions}}\label{subs:3.1}
The secondary masses and the periods of WD + MS systems at the
moment of explosion are basic input parameters when one simulates
the interaction between supernova ejecta and its companion. We
show the distributions of the masses and the periods for different
metallicities $Z$ in Figs. \ref{permms0302} to
\ref{permms00010001}, where $M_{\rm WD}=1.378 M_{\odot}$. Here, we
only show the cases of $\alpha_{\rm CE}=1.0$ except for $Z=0.0001$
(see the bottom panel in Fig. \ref{permms00010001}). For other
$Z$, $\alpha_{\rm CE}$ does not significantly affect the final
distributions.

From these figures, we see that generally, the companions with a
high metallicity $Z$ have larger final masses, and roughly when
$Z>0.004$, the final periods center on 1 to 1.6 days, smaller than
those of the cases of $Z\leq0.004$ (centering on 2.5 to 6.5 days).
These phenomena are mainly derived from the distributions of the
initial parameters of WD +MS systems with various $Z$. Paper I has
shown that a high metallicity leads to a more massive initial
secondary mass(see Figs. 4 and 10 in Paper I), and the systems
with $Z>0.004$ have a shorter initial period than those with
$Z\leq0.004$ (see Fig. 11 in Paper I). Meanwhile, metallicity also
affects mass transfer process. A high metallicity results in a
lower mass transfer rate (\citealt{LAN00}), and then less material
loses from binary system as optically thick wind
(\citealt{HAC96}), which means that more transferred materials are
accumulated on the WD. Therefore, mass-ratio inversion for a WD
+ MS system with a high metallicity is delayed, and then the WD +
MS system is more likely to decrease its orbital period, which may lead to a relatively low
period (see also Paper I and \citealt{MENGYG09}).

The bottom panel in Fig. \ref{permms00010001} shows the case for
$Z=0.0001$ and $\alpha_{\rm CE}=3.0$. In the figure, there are two
``islands" in the ($M_{\rm 2}^{\rm SN}, \log P_{\rm SN}$) plane,
i.e. a low-mass low-period one and a high-mass high-period one,
which is significantly different from the top panel in Fig.
\ref{permms00010001}. We have not found the two islands in other
cases. As shown in subsection \ref{subs:2.2} and paper I, there
are three channels leading to close WD + MS binary systems, i.e.
He star channel, EAGB channel and TPAGB channel. Metallicity and
$\alpha_{\rm CE}$ may systemically affect the channels (see Paper
I for the influence of Metallicity and $\alpha_{\rm CE}$ on the
channels for details). A low metallicity and a high $\alpha_{\rm
CE}$, i.e. $Z=0.0001$ and $\alpha_{\rm CE}=3.0$, may result in
double peaks in the distributions of the initial masses of CO WDs
and the initial periods of WD + MS systems leading to SNe Ia (see
Figs. 9 and 11 in Paper I). The low-mass peak and the low-period
peak are from the EAGB channel, while the high-mass peak and the
high-period peak are from the TPAGB channel. It is obvious that
the low-mass peak of CO WDs and the low-period peak may lead to
the low-mass low-period island in the bottom panel in Fig.
\ref{permms00010001} since more materials are needed to transfer
onto CO WD from mass donor if a low-mass WD is adopted. For a
similar reason, the high-mass high-period island in the bottom
panel in Fig. \ref{permms00010001} results from the high-mass peak
and high-period peak from TPAGB channel.

When $Z>0.0001$ and $\alpha_{\rm CE}=3.0$, no WD + MS systems are
from TPAGB channel, which results in the disappearance of double
peaks (see \citealt{MENGCH09} for details). This can naturally
explain why the double islands in Fig. \ref{permms0302} to the top
panel in Fig. \ref{permms00010001} disappear.

These figures may also help to check whether some binary system
observed can explode as SNe Ia or not. In these figures, a
recurrent nova, U Sco, is indicated by filled square
(\citealt{SR95}; \citealt{HK00a, HK00b}). The WD mass of U Sco is
about $1.37 M_{\odot}$, and its companion is a MS sequence star of
$1.5 M_{\odot}$ (\citealt{HK00a, HK00b}). The orbital period is
1.23056 days (\citealt{SR95}). \citet{HK00a, HK00b} studied the
system carefully and concluded that the WD mass can grow until an
SN Ia explosion is triggered after $\sim10^{\rm 5}$ yr. If the WD
of U Sco may explode as a SN Ia finally, the companion should have
a slightly smaller mass. Its period will decrease firstly, and
then may increase if mass-ratio reverses. Then, its final position
in ($M_{\rm 2}^{\rm SN}, \log P_{\rm SN}$) plane will move to a
lower mass from its present position at least, and may enters into
the most probable area (see the bottom panel in Fig
\ref{permms0302}). From its present position in ($M_{\rm 2}^{\rm
SN}, \log P_{\rm SN}$) plane, it is very likely for U Sco to
explode as a SN Ia (see also \citealt{HKN08}; \citealt{MENGYG09}).

\begin{figure}
 \centering
\centerline{\psfig{figure=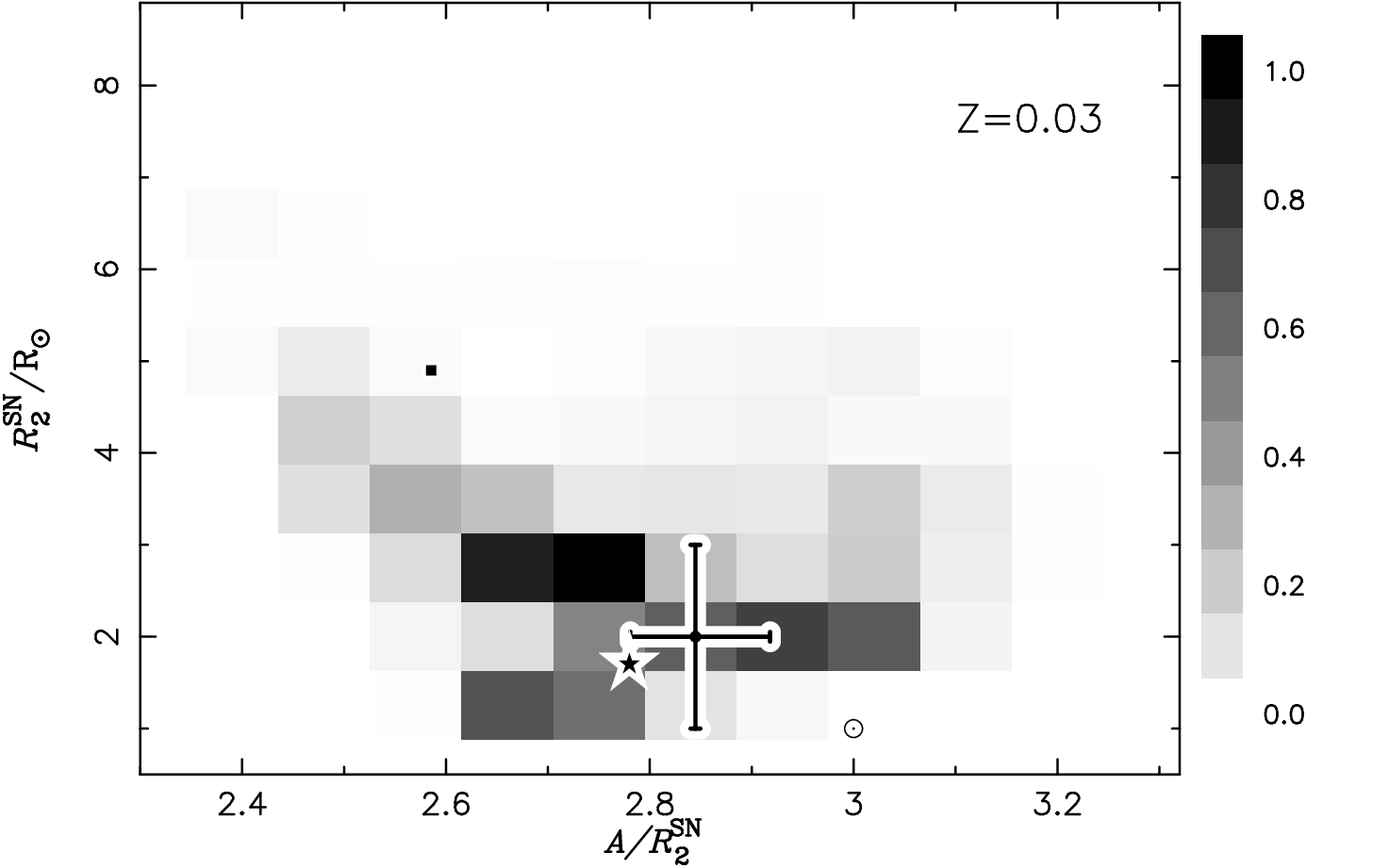,width=0.3\textheight,height=0.40\textwidth,bbllx=60pt,bblly=100pt,bburx=585pt,bbury=676pt,angle=270}}
\centerline{\psfig{figure=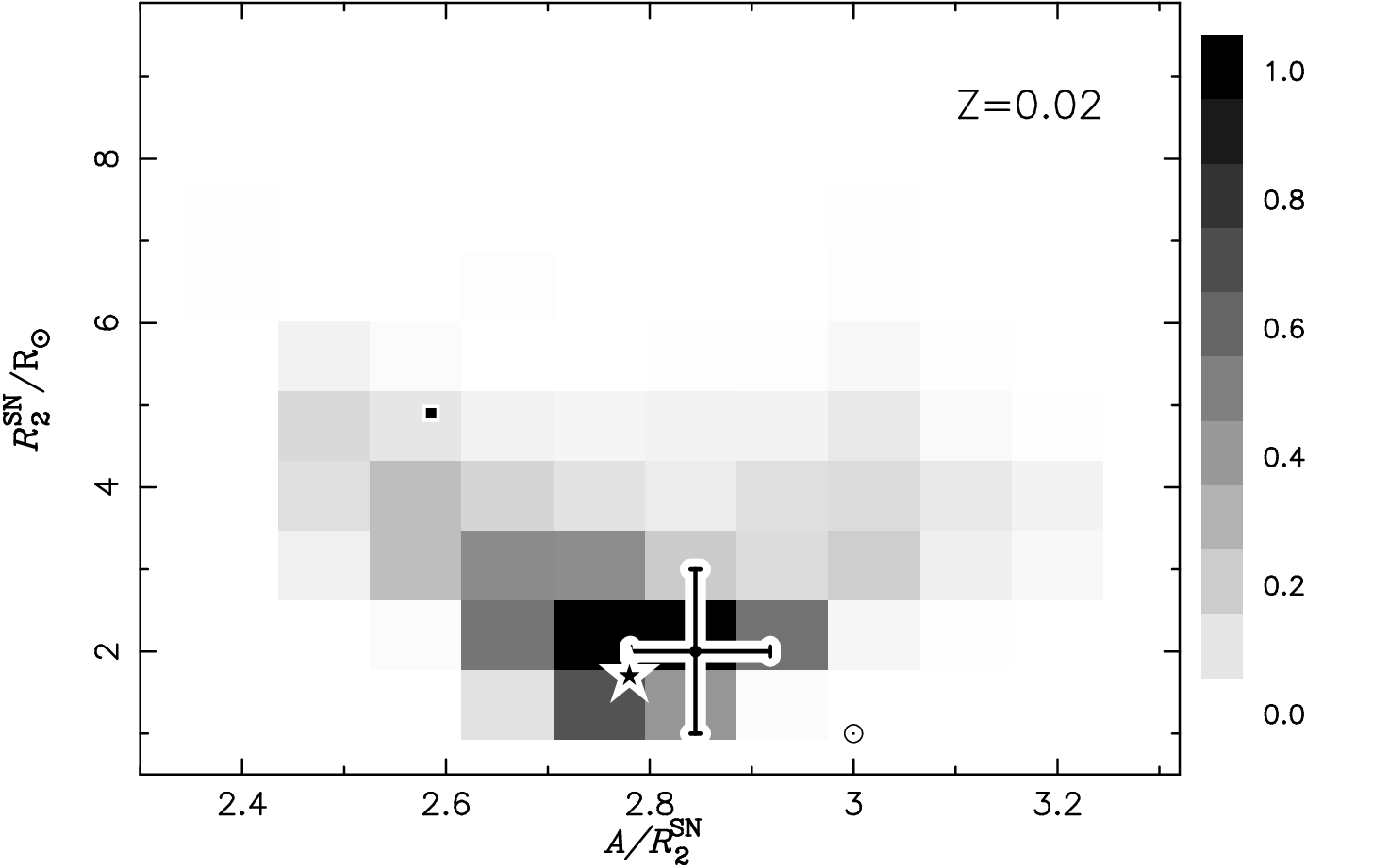,width=0.3\textheight,height=0.40\textwidth,bbllx=60pt,bblly=100pt,bburx=585pt,bbury=676pt,angle=270}}
 \suppressfloats[t]
 \caption{The distribution of the radii and the ratios of
separations to radii of companions at the moment of supernova
explosion for the case of $\alpha_{\rm CE}=1.0$. Cross represents
Tycho G, which is a potential candidate of the companion of
Tycho's supernova (\citealt{RUI04}; \citealt{BRA04}). The length
of the cross represents observational error. Solar symbol and
\textbf{star} represent the main-sequence and subgiant companion
model, respectively, which are applied to simulate the collision
between supernova ejecta and its companion in \citet{MAR00}. The
position of a recurrent nova, U Sco, is indicated by the filled
square (\citealt{SR95}; \citealt{HK00a, HK00b}). Top: $Z=0.03$;
Bottom: $Z=0.02$.}
 \label{ratrsn0302}
\end{figure}

\begin{figure}
 \centering
\centerline{\psfig{figure=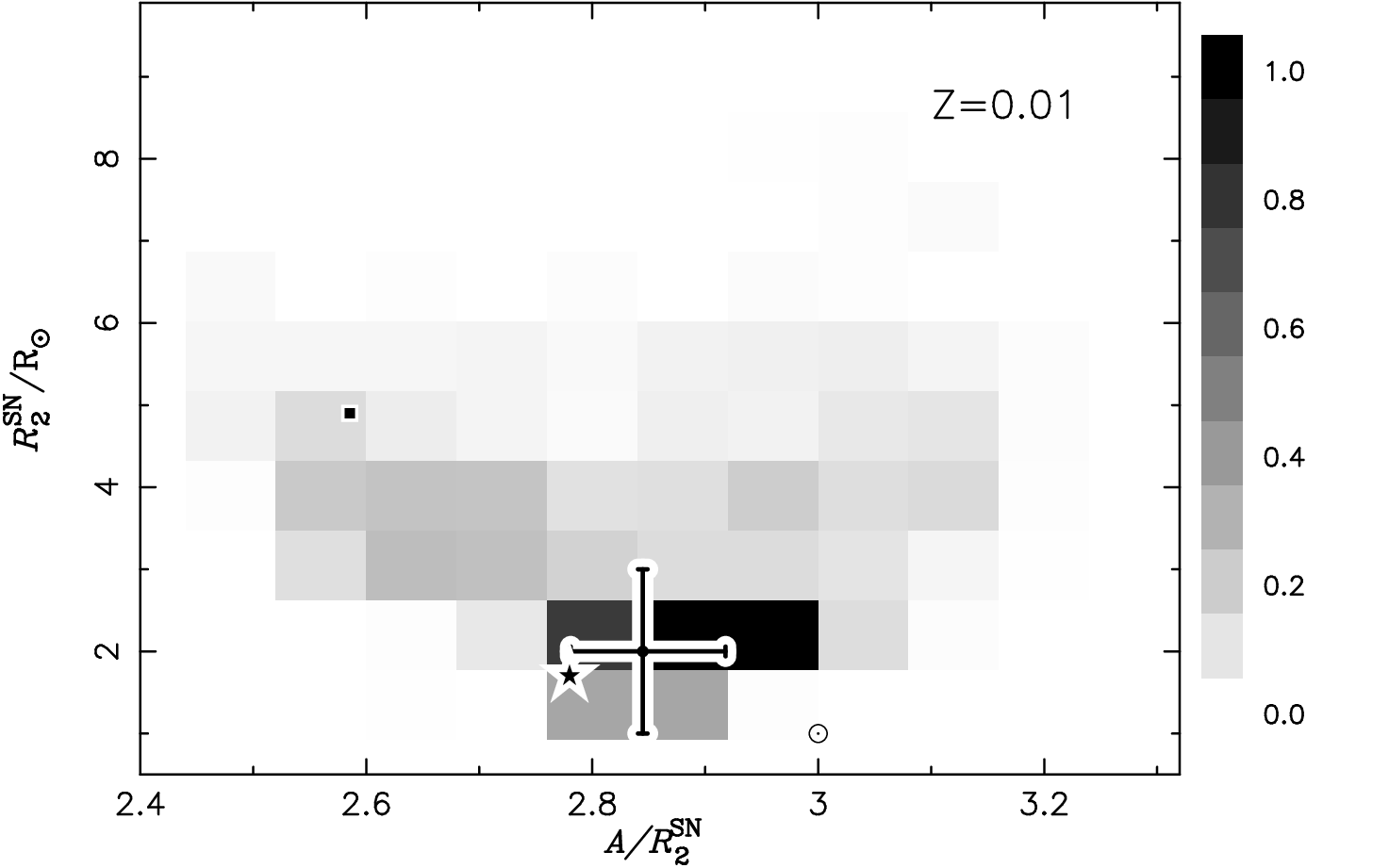,width=0.3\textheight,height=0.40\textwidth,bbllx=60pt,bblly=100pt,bburx=585pt,bbury=676pt,angle=270}}
\centerline{\psfig{figure=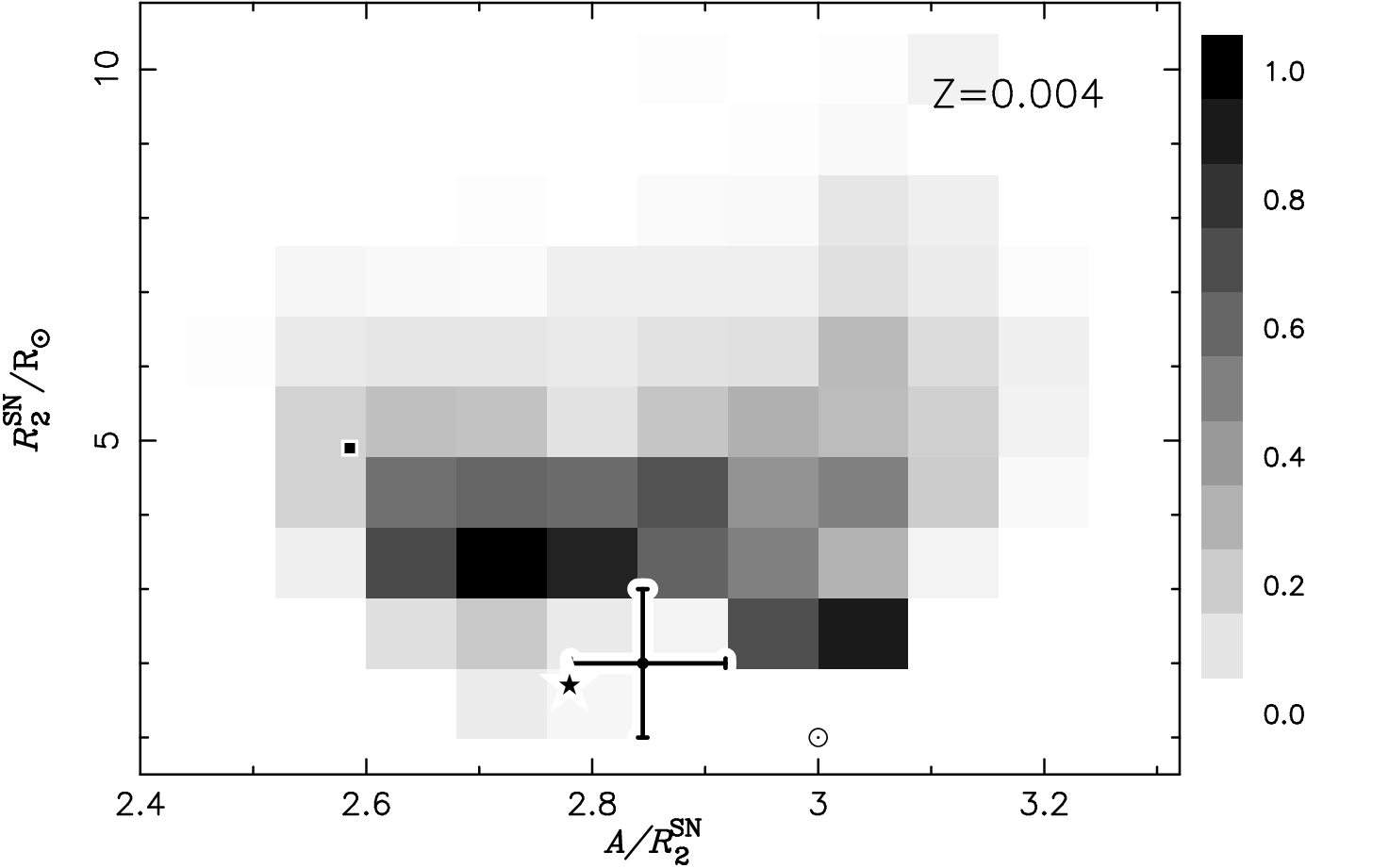,width=0.3\textheight,height=0.40\textwidth,bbllx=60pt,bblly=100pt,bburx=585pt,bbury=676pt,angle=270}}
 \suppressfloats[t]
 \caption{Similar to Fig \ref{ratrsn0302} but for $Z=0.01$ (Top) and $Z=0.004$ (Bottom).}
 \label{ratrsn01004}
\end{figure}

\begin{figure}
 \centering
\centerline{\psfig{figure=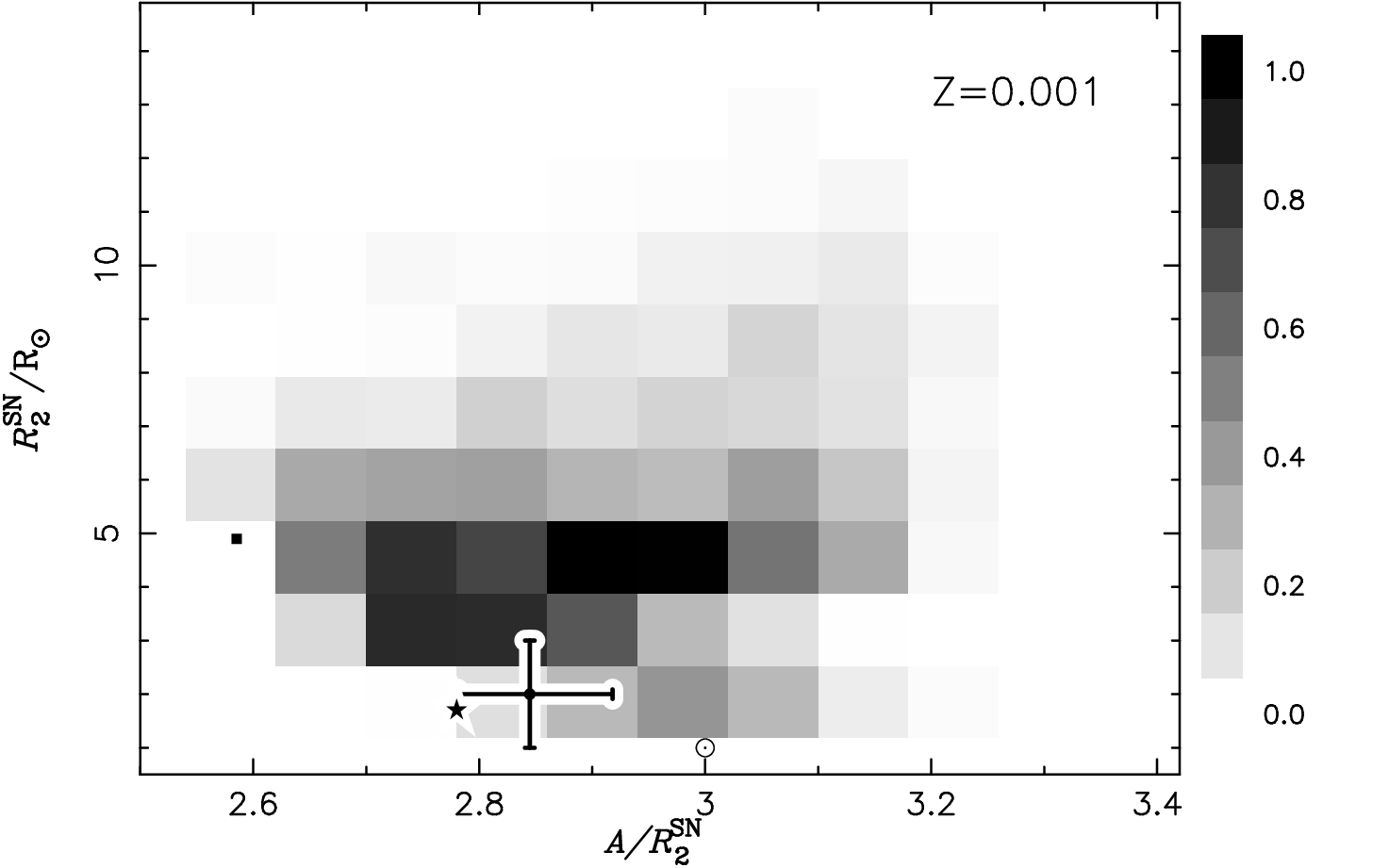,width=0.3\textheight,height=0.40\textwidth,bbllx=60pt,bblly=100pt,bburx=585pt,bbury=676pt,angle=270}}
\centerline{\psfig{figure=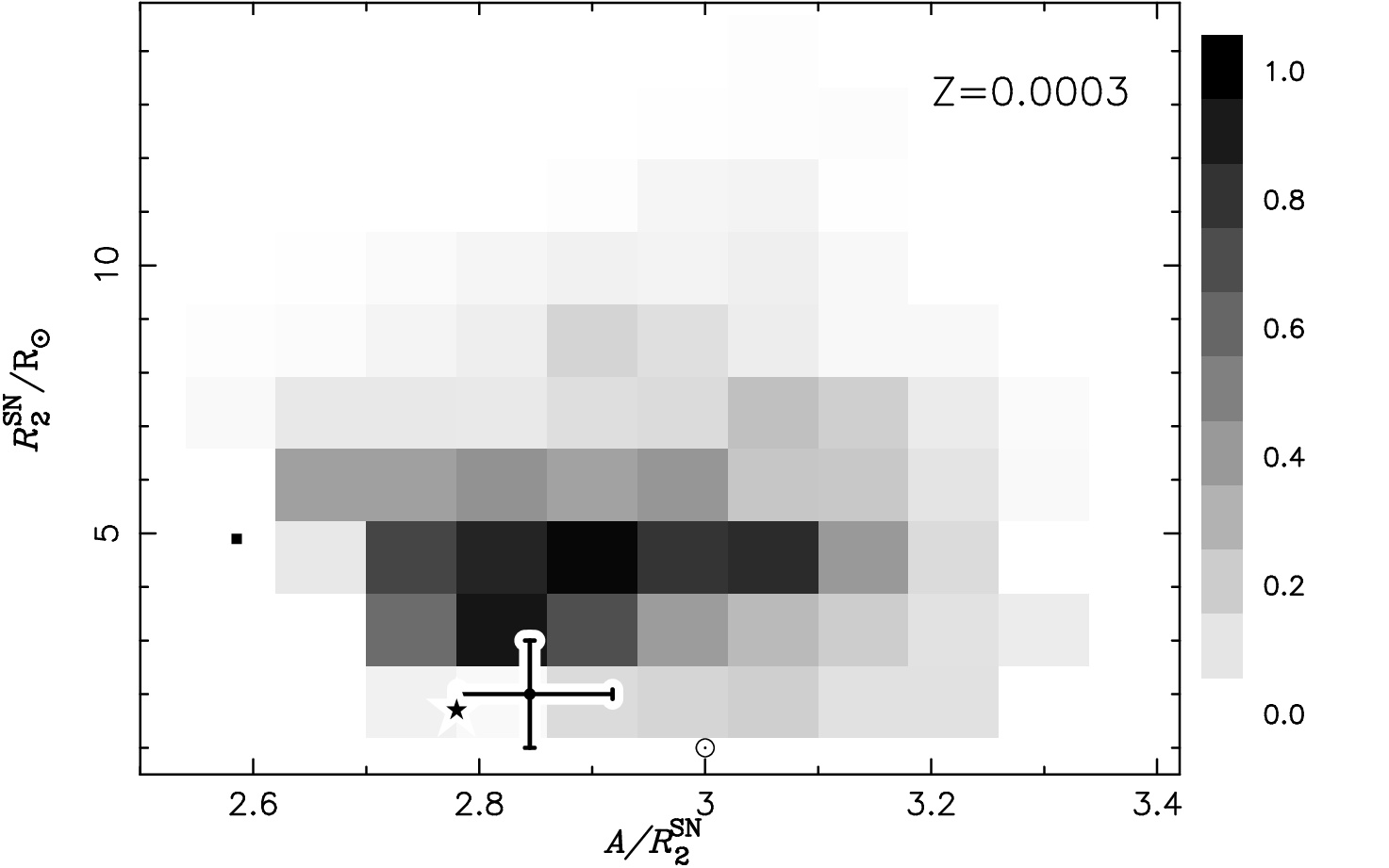,width=0.3\textheight,height=0.40\textwidth,bbllx=60pt,bblly=100pt,bburx=585pt,bbury=676pt,angle=270}}
 \suppressfloats[t]
 \caption{Similar to Fig \ref{ratrsn0302} but for $Z=0.001$ (Top) and $Z=0.0003$ (Bottom).}
 \label{ratrsn0010003}
\end{figure}

\begin{figure}
 \centering
\centerline{\psfig{figure=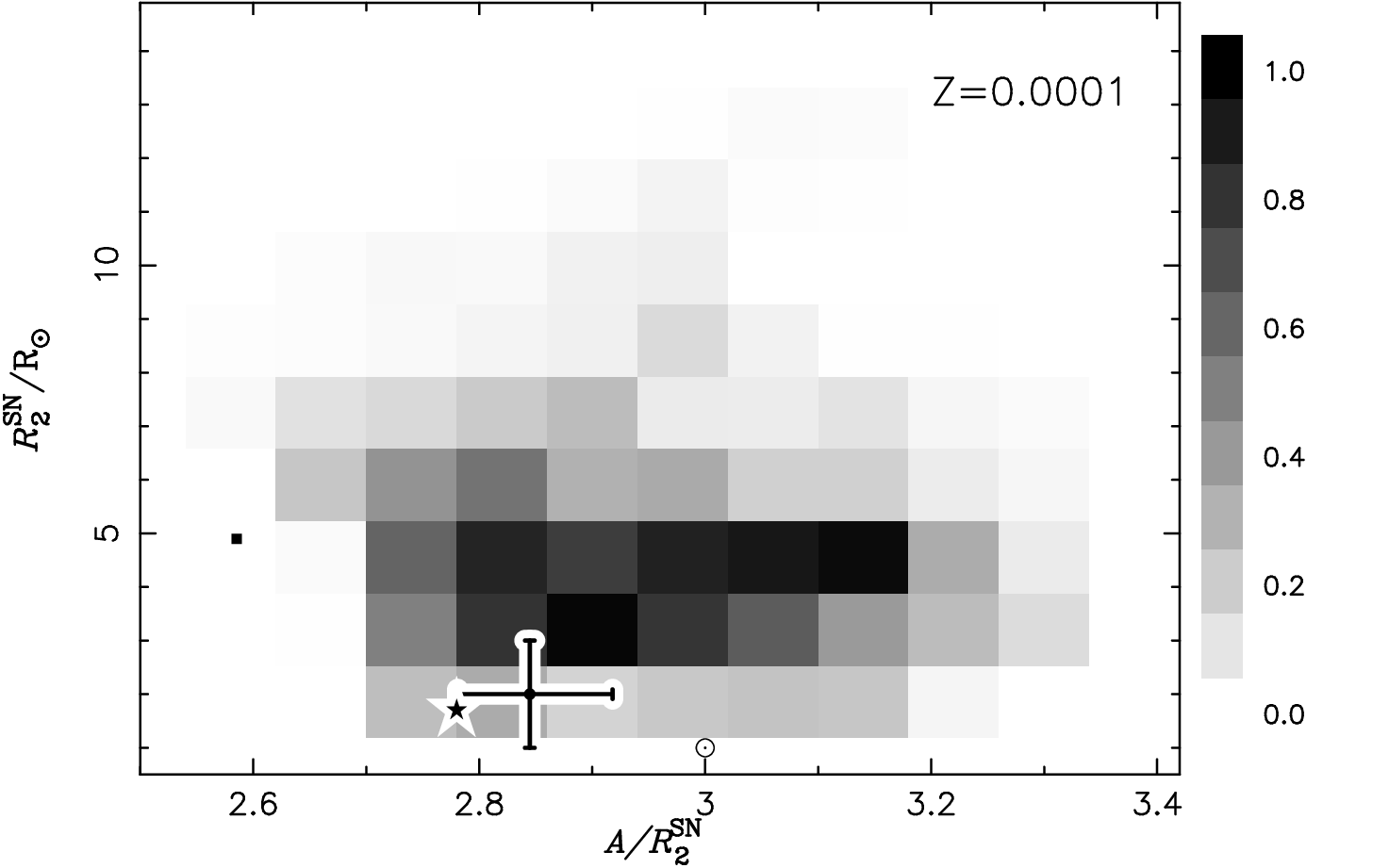,width=0.3\textheight,height=0.40\textwidth,bbllx=60pt,bblly=100pt,bburx=585pt,bbury=676pt,angle=270}}
\centerline{\psfig{figure=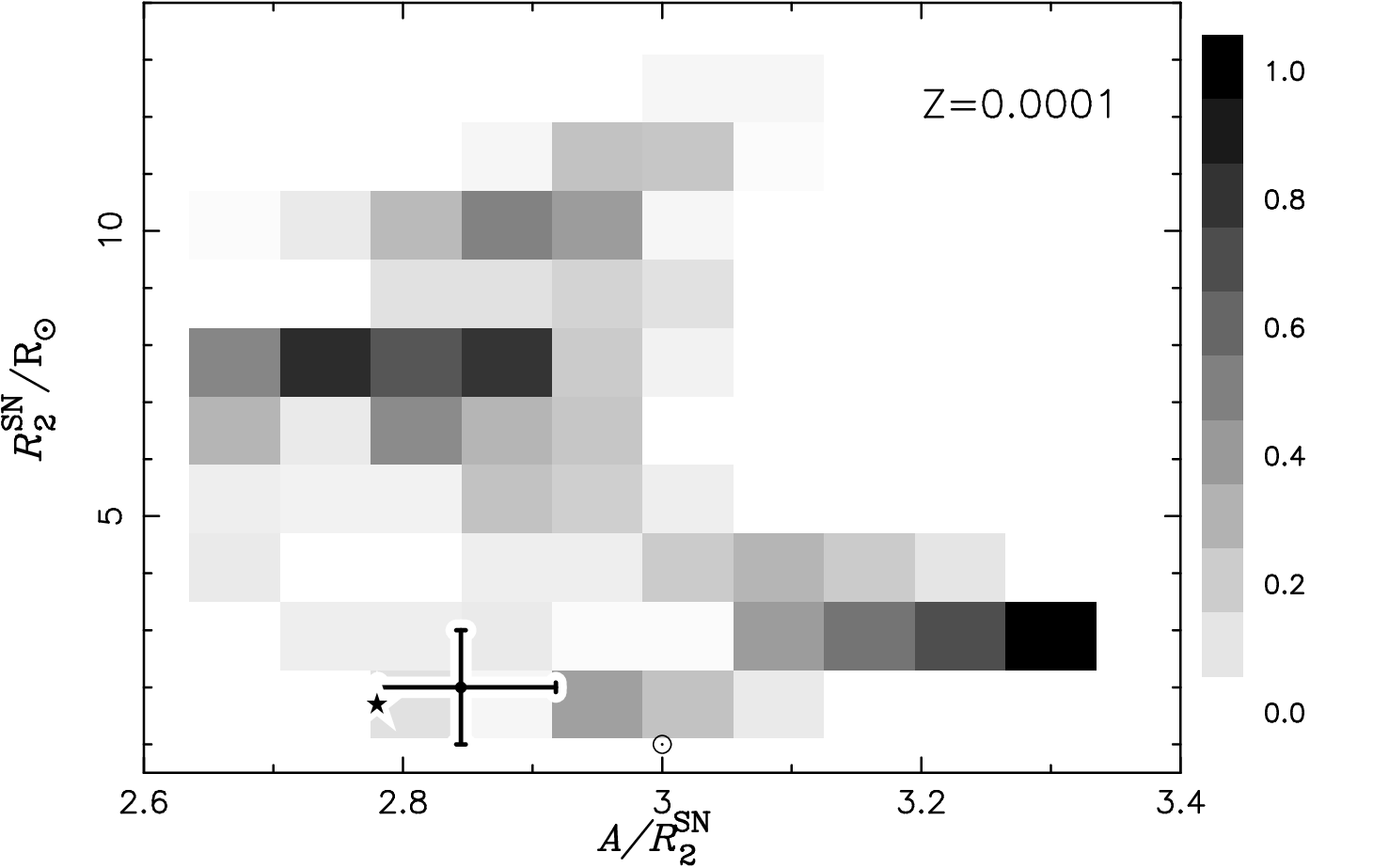,width=0.3\textheight,height=0.40\textwidth,bbllx=60pt,bblly=100pt,bburx=585pt,bbury=676pt,angle=270}}
 \suppressfloats[t]
 \caption{Similar to Fig \ref{ratrsn0302} but for $Z=0.0001$. Top: $\alpha_{\rm
CE}=1.0$; Bottom: $\alpha_{\rm CE}=3.0$.}
 \label{ratrsn00010001}
\end{figure}

\subsection{Radii and the ratios of separations to radii of companions at the moment of explosions}\label{subs:3.2}
The radius of a companion, $R_{\rm 2}^{\rm SN}$, and the ratio of
separation to the  companion's radius, $A/R_{\rm 2}^{\rm SN}$, at
the moment of explosion are important parameters for simulating
the interaction between supernova ejecta and the companion
(\citealt{MAR00}). There is even a linear dependence of kick
velocity and the mass of stripped material on $\log(A/R_{\rm
2}^{\rm SN})$ (\citealt{MAR00}; \citealt{MENGCH07}). After the
interaction, the companion reestablishes dynamical equilibrium
quickly while it is still in a process into thermal equilibrium.
The process into thermal equilibrium may last for $10^{\rm
3}$ - $10^{\rm 4}$ yr. It is similar that a companion star is also not
back into thermal equilibrium at the moment of supernova explosion
since mass transfer is processing before supernova explosion. So,
considering the amount of the stripped material from companions is
small (see section \ref{sect:1} or next subsection,
\citealt{MAT05}; \citealt{LEO07}; \citealt{MENGCH07}), the radius
of a companion at the moment of supernova explosion may represent
its radius after the impact of a SN Ia to some extent, and then
our results can also compare with observation directly. Under the
assumption that the radius of a companion is equal to its Roche
lobe radius (\citealt{EGG83}), we show the distributions of
$R_{\rm 2}^{\rm SN}$ and $A/R_{\rm 2}^{\rm SN}$ at the moment of
explosion for various $Z$ and $\alpha_{\rm CE}=1.0$ in Figs.
\ref{ratrsn0302} to \ref{ratrsn00010001}. The cases with
$\alpha_{\rm CE}=3.0$ are similar to that with $\alpha_{\rm
CE}=1.0$ except for $Z=0.0001$ (the bottom panel in in
Fig \ref{ratrsn00010001}).

The age of Tycho's supenrova
is about 440 yr, which means that the suggested companion star of
Tycho's supenrova, Tycho G, is still not back into thermal equilibrium
and its radius at present might be similar to that at the moment of supernova explosion, also similar to its Roche lobe radius.
Assuming that the radius of Tycho G at present is equal to its Roche lobe radius
at the moment of supernova explosion,
and using the equation in \citet{EGG83}, we may obtain the $A/R_{\rm 2}^{\rm SN}$ of Tycho G.
The cross in the figures represents the Tycho G
(\citealt{RUI04}; \citealt{BRA04}). The solar symbol and open
pentacle denote the MS (a solar model) and SG companion model,
which are applied by \citet{MAR00} to simulate the interaction
between supernova ejecta and its companion. Generally, the radius
of a companion decreases with metallicity, which is directly
originated from the dependence of the periods of systems on
metallicity (see last subsection).

We see in these figures that Tycho G is consistent with our
high-metallicity results. The SG companion model used in
\citet{MAR00} also matches with our high-metallicity results since
this model is obtained from \citet{LI97}, who used a method
and a stellar evolution code similar to that in Paper I with $Z=0.02$.
However, the MS model in \citet{MAR00} is obviously departed from
the ($R_{\rm 2}^{\rm SN}, A/R_{\rm 2}^{\rm SN}$) plane for all
metallicities, which implies that a solar model is not a
reasonable model when simulating the interaction between supernova
ejecta and its companion.

There are also two islands in the ($R_{\rm 2}^{\rm SN},
A/R_{\rm 2}^{\rm SN}$) plane with $Z=0.0001$ and $\alpha_{\rm
CE}=3.0$ (see the bottom panel in Fig \ref{ratrsn00010001}). The
reason is same to that interpreted in subsections \ref{subs:3.1}.

We also showed the recurrent nova, U Sco, in these figures. Since
the orbital period of the system will reduce, the radius of the MS
companion in the system should decrease, and $A/R_{\rm 2}^{\rm
SN}$ increase based on the equation of \citet{EGG83}. The position
of U Sco will move towards right-lower region in ($R_{\rm 2}^{\rm
SN}, A/R_{\rm 2}^{\rm SN}$) plane, and may enter into the most
probable region (see the bottom panel in Fig. \ref{ratrsn0302}).
The bottom panel in Fig. \ref{ratrsn0302} also shows that the
recurrent nova, U Sco, probably explodes as a SN Ia.

\begin{figure}
 \centering
\centerline{\psfig{figure=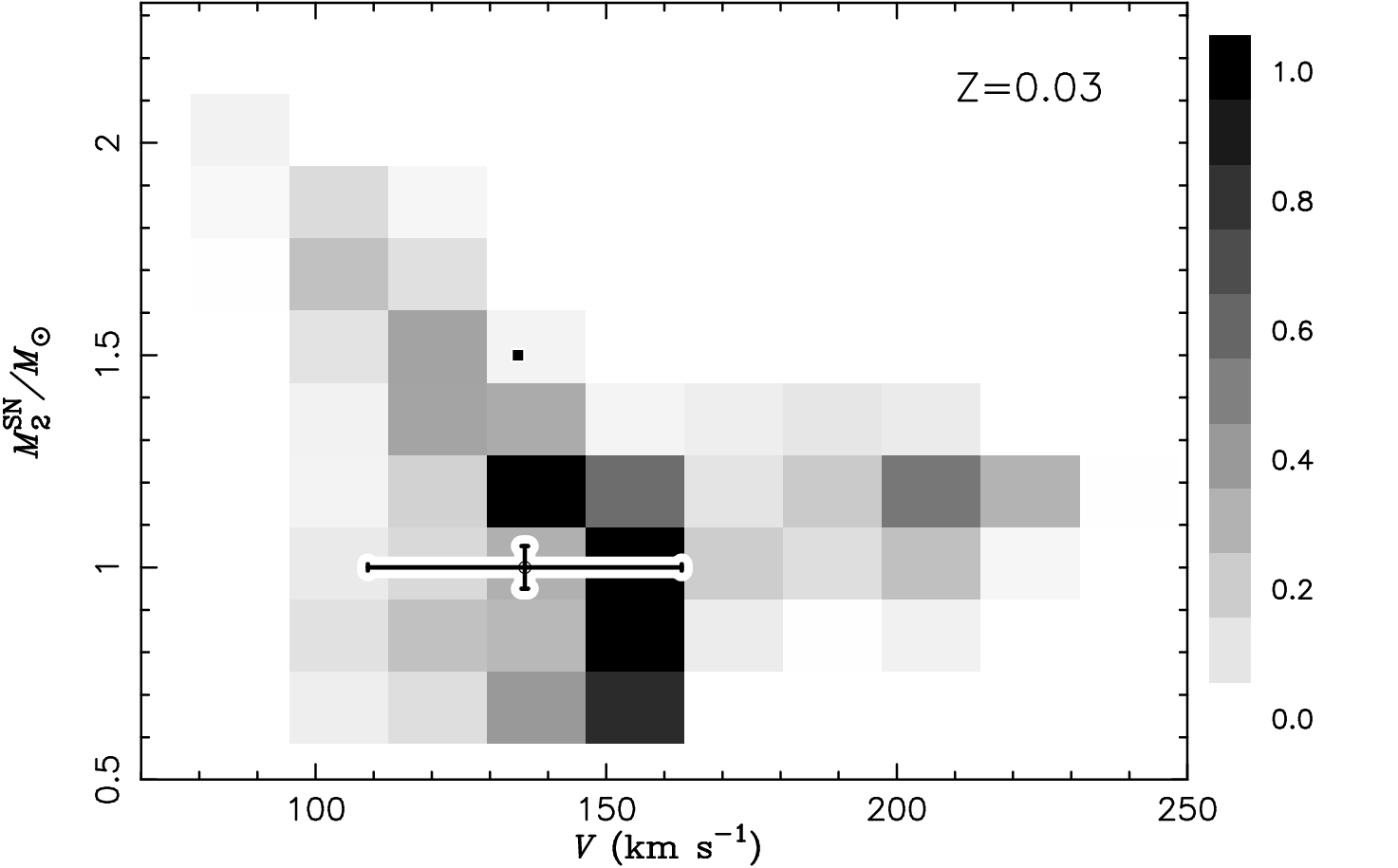,width=0.3\textheight,height=0.40\textwidth,bbllx=60pt,bblly=100pt,bburx=585pt,bbury=676pt,angle=270}}
\centerline{\psfig{figure=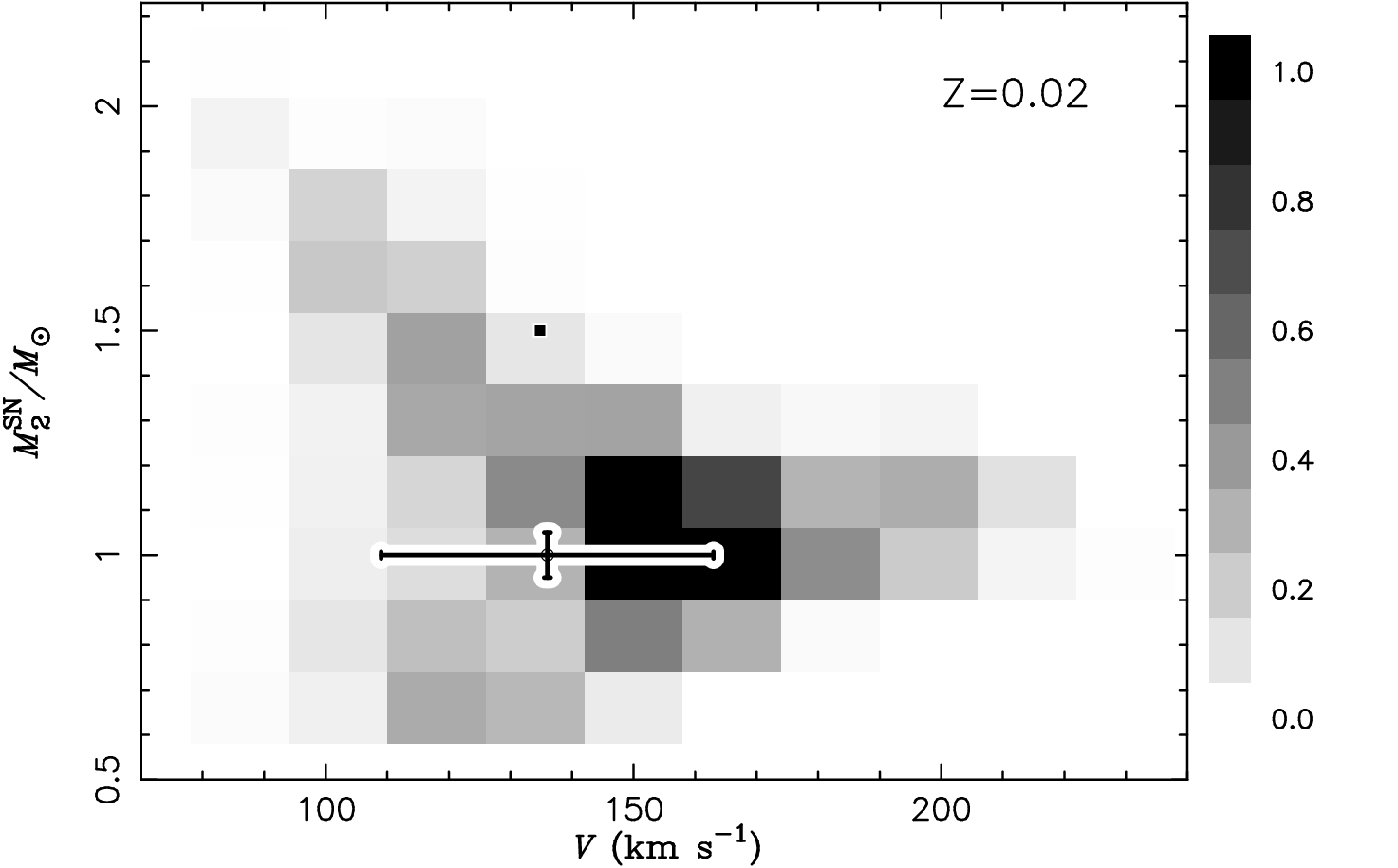,width=0.3\textheight,height=0.40\textwidth,bbllx=60pt,bblly=100pt,bburx=585pt,bbury=676pt,angle=270}}
 \suppressfloats[t]
 \caption{The distribution of the
masses and the space velocities of companions in SNe Ia remnant
for $Z=0.03$ and $\alpha_{\rm CE}=1.0$. The cross represents the
position of Tycho G, which is the potential candidate of the
companion of Tycho's supernova (\citealt{RUI04}; \citealt{BRA04}),
and the length of the bars of the cross represents observational
error. The position of the MS star of a recurrent nova, U Sco, is indicated by
the filled square, where its space velocity is its orbital velocity relative to the mass center of the system (\citealt{SR95}; \citealt{HK00a, HK00b}). Top:
$Z=0.03$; Bottom: $Z=0.02$.}
 \label{vorbmms0302}
\end{figure}

\begin{figure}
 \centering
\centerline{\psfig{figure=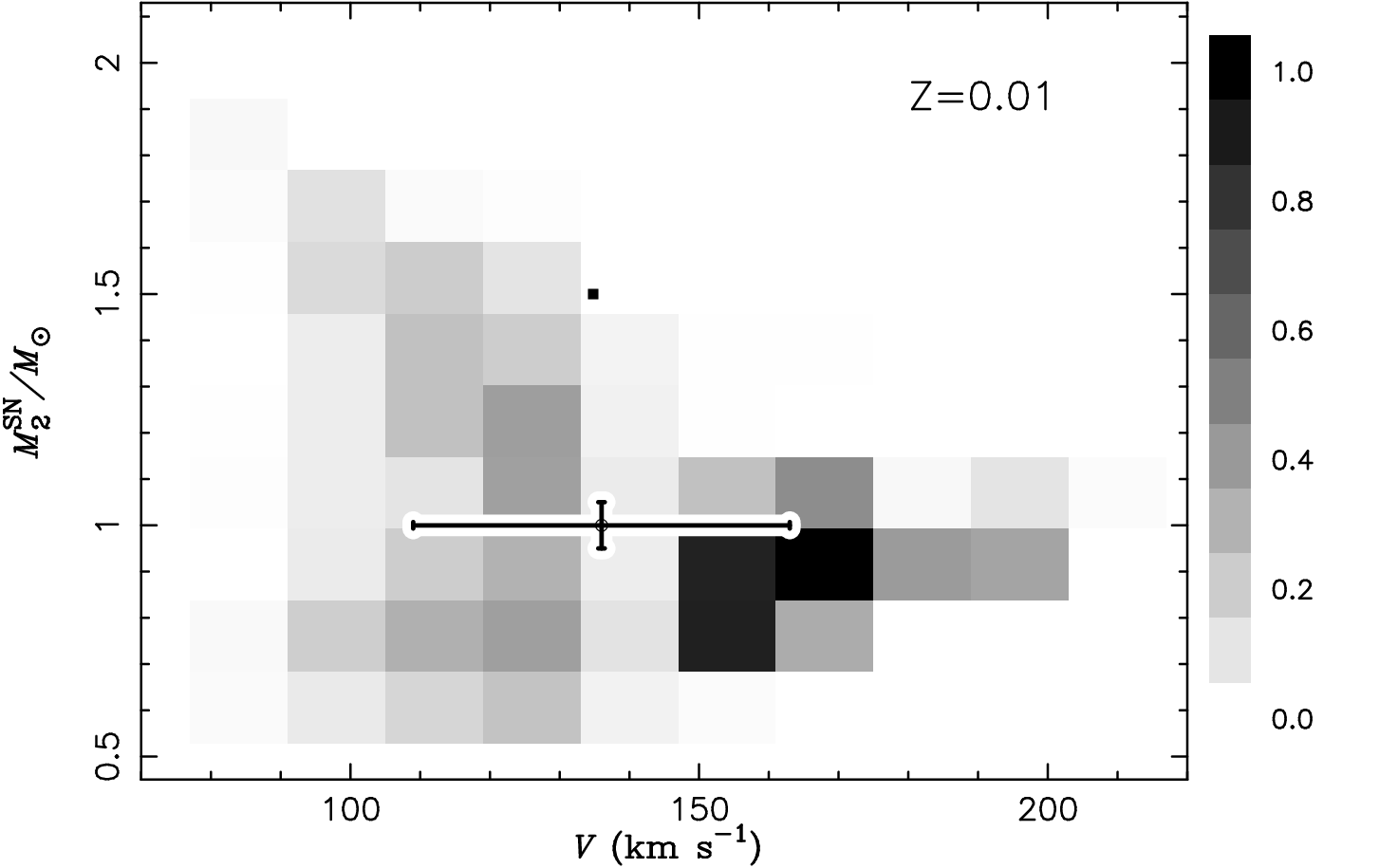,width=0.3\textheight,height=0.40\textwidth,bbllx=60pt,bblly=100pt,bburx=585pt,bbury=676pt,angle=270}}
\centerline{\psfig{figure=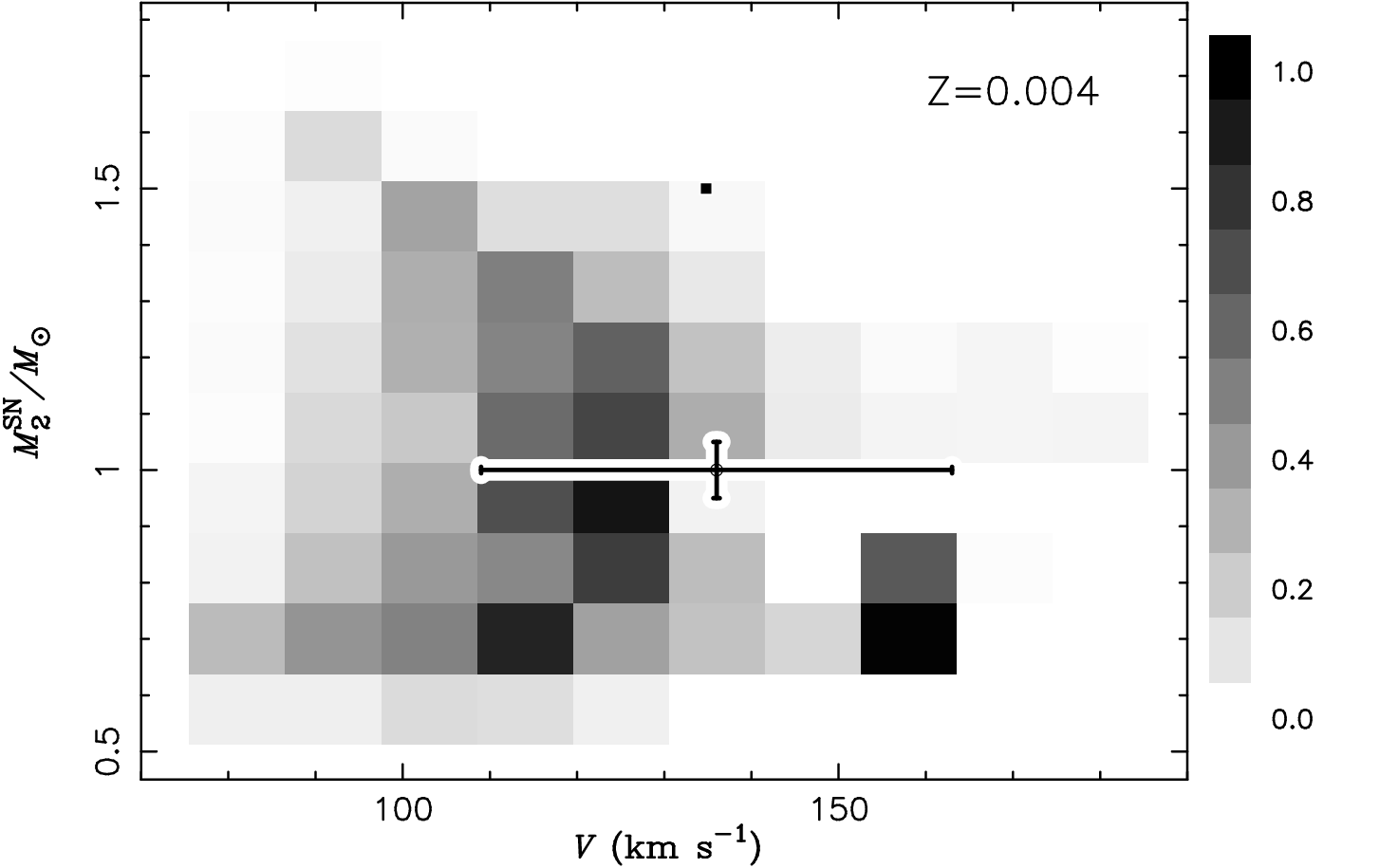,width=0.3\textheight,height=0.40\textwidth,bbllx=60pt,bblly=100pt,bburx=585pt,bbury=676pt,angle=270}}
 \suppressfloats[t]
 \caption{Similar to Fig \ref{vorbmms0302} but for $Z=0.01$ (Top) and $Z=0.004$ (Bottom).}
 \label{vorbmms01004}
\end{figure}

\begin{figure}
 \centering
\centerline{\psfig{figure=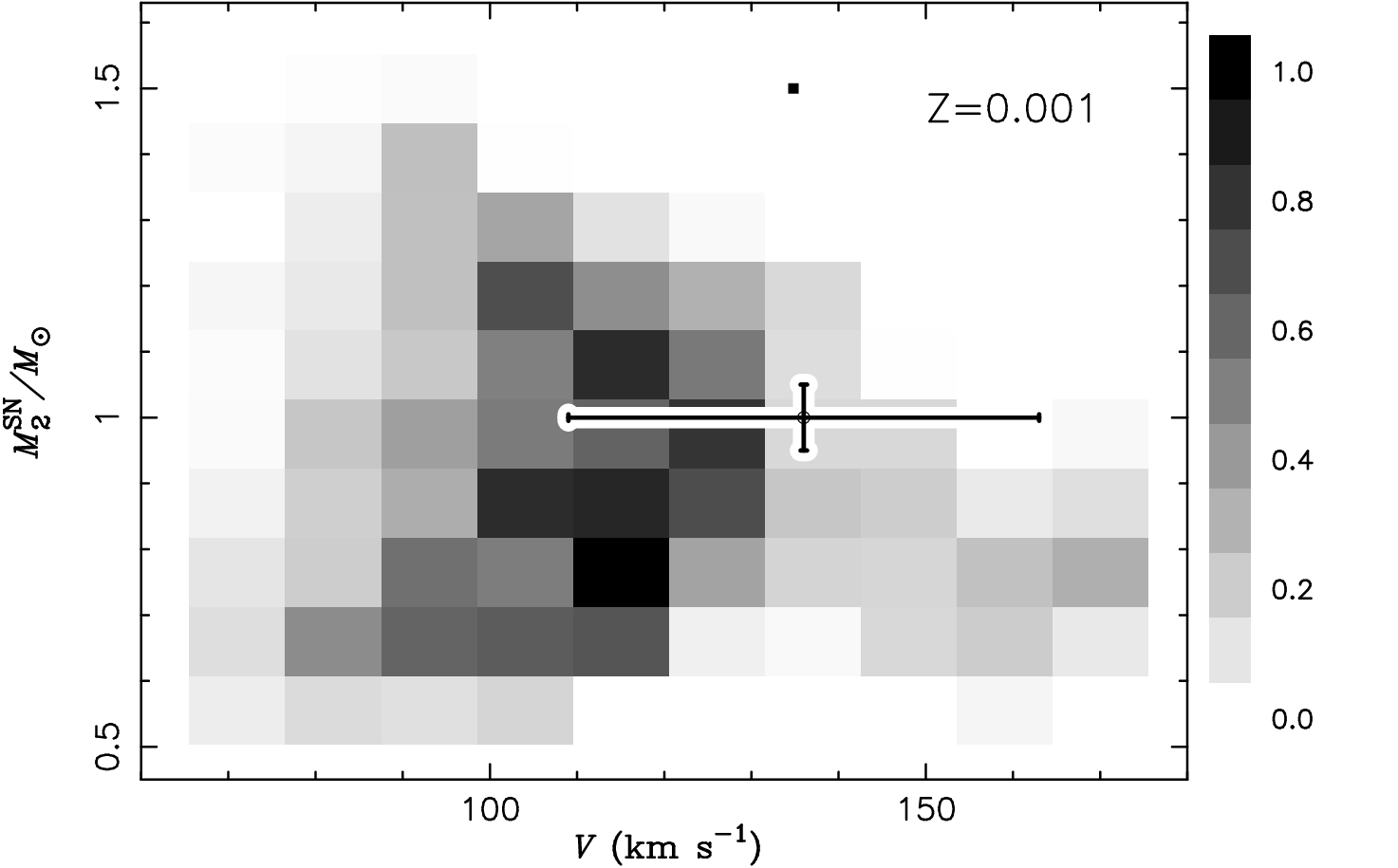,width=0.3\textheight,height=0.40\textwidth,bbllx=60pt,bblly=100pt,bburx=585pt,bbury=676pt,angle=270}}
\centerline{\psfig{figure=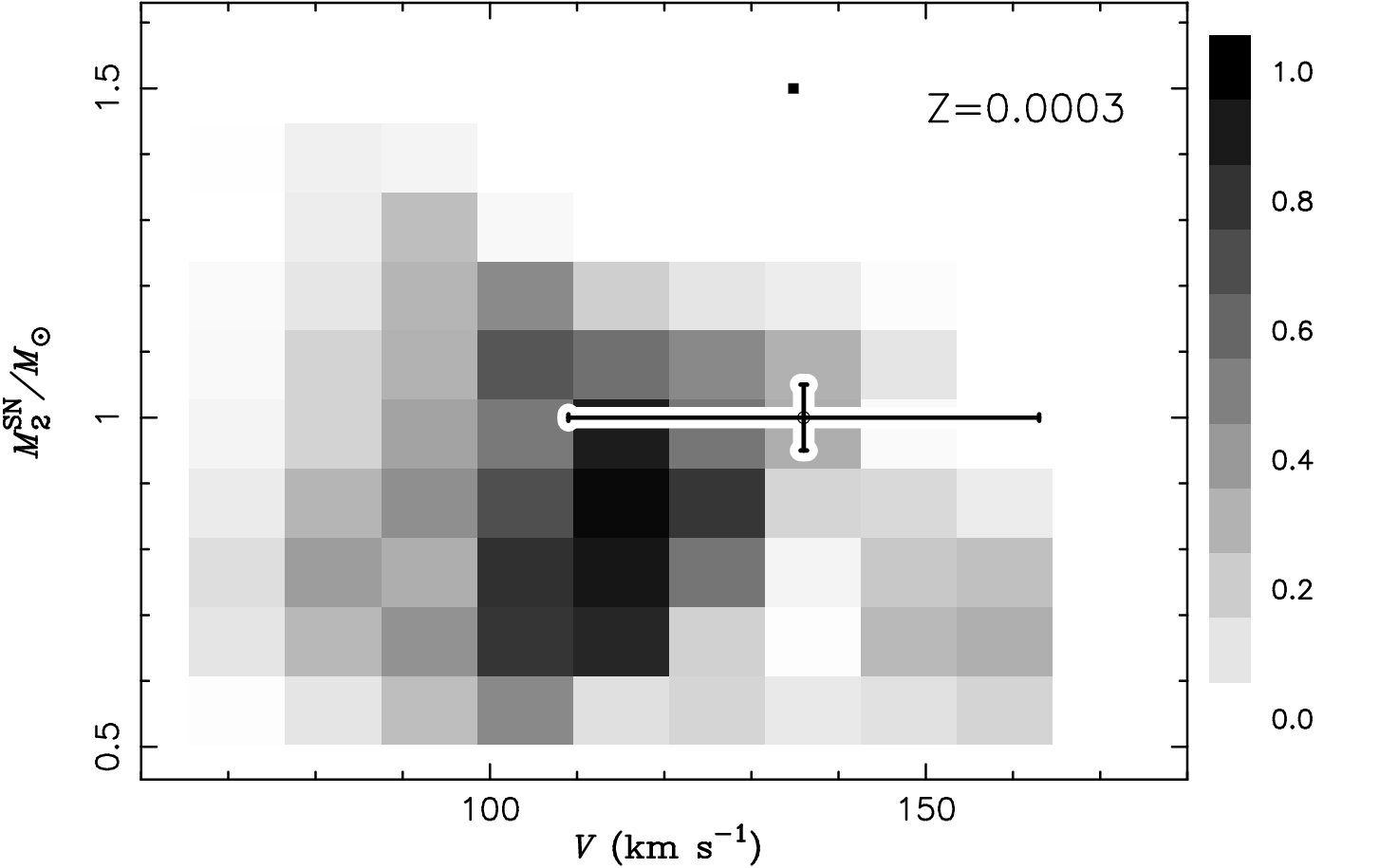,width=0.3\textheight,height=0.40\textwidth,bbllx=60pt,bblly=100pt,bburx=585pt,bbury=676pt,angle=270}}
 \suppressfloats[t]
 \caption{Similar to Fig \ref{vorbmms0302} but for $Z=0.001$ (Top) and $Z=0.0003$ (Bottom).}
 \label{vorbmms0010003}
\end{figure}

\begin{figure}
 \centering
\centerline{\psfig{figure=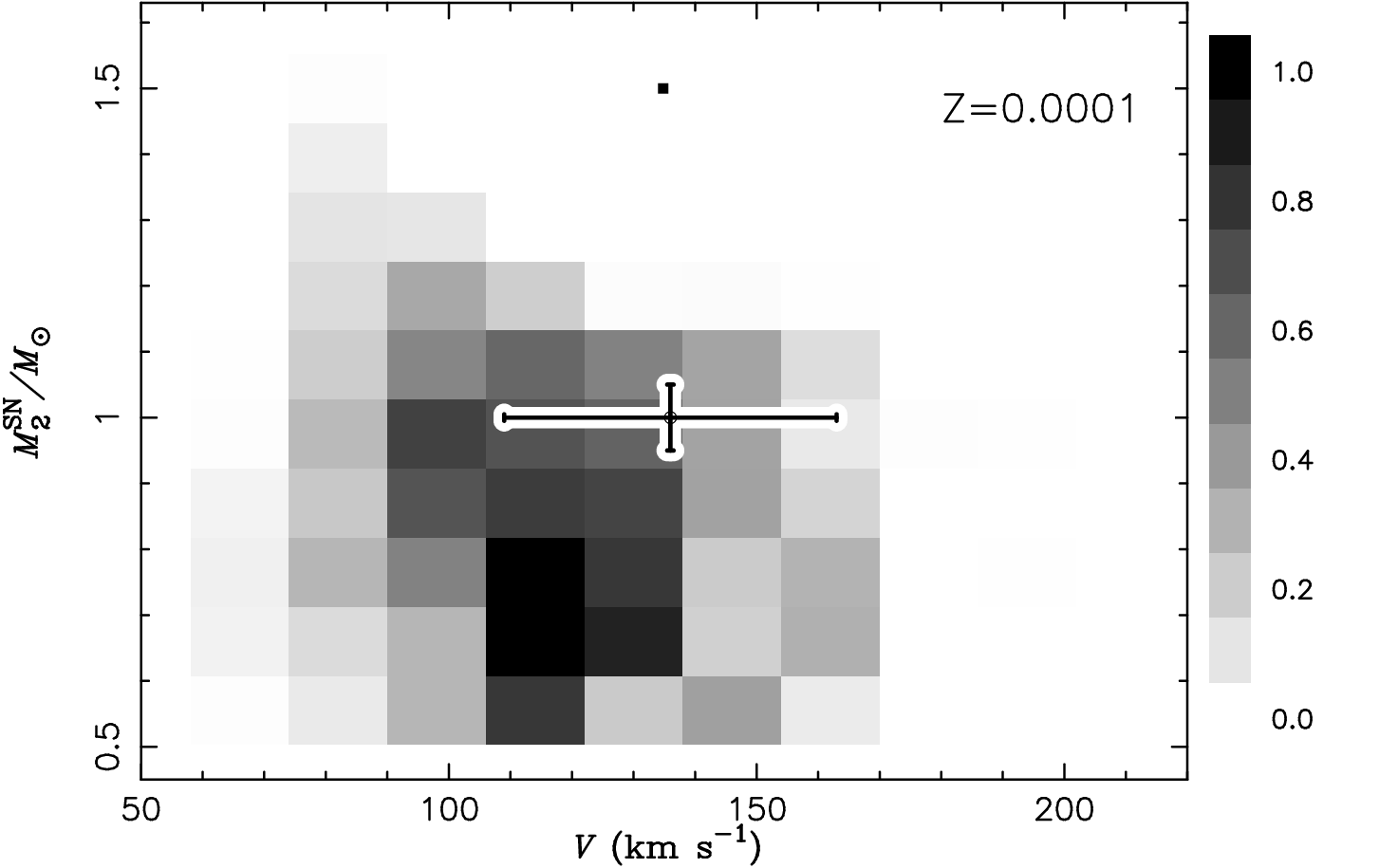,width=0.3\textheight,height=0.40\textwidth,bbllx=60pt,bblly=100pt,bburx=585pt,bbury=676pt,angle=270}}
\centerline{\psfig{figure=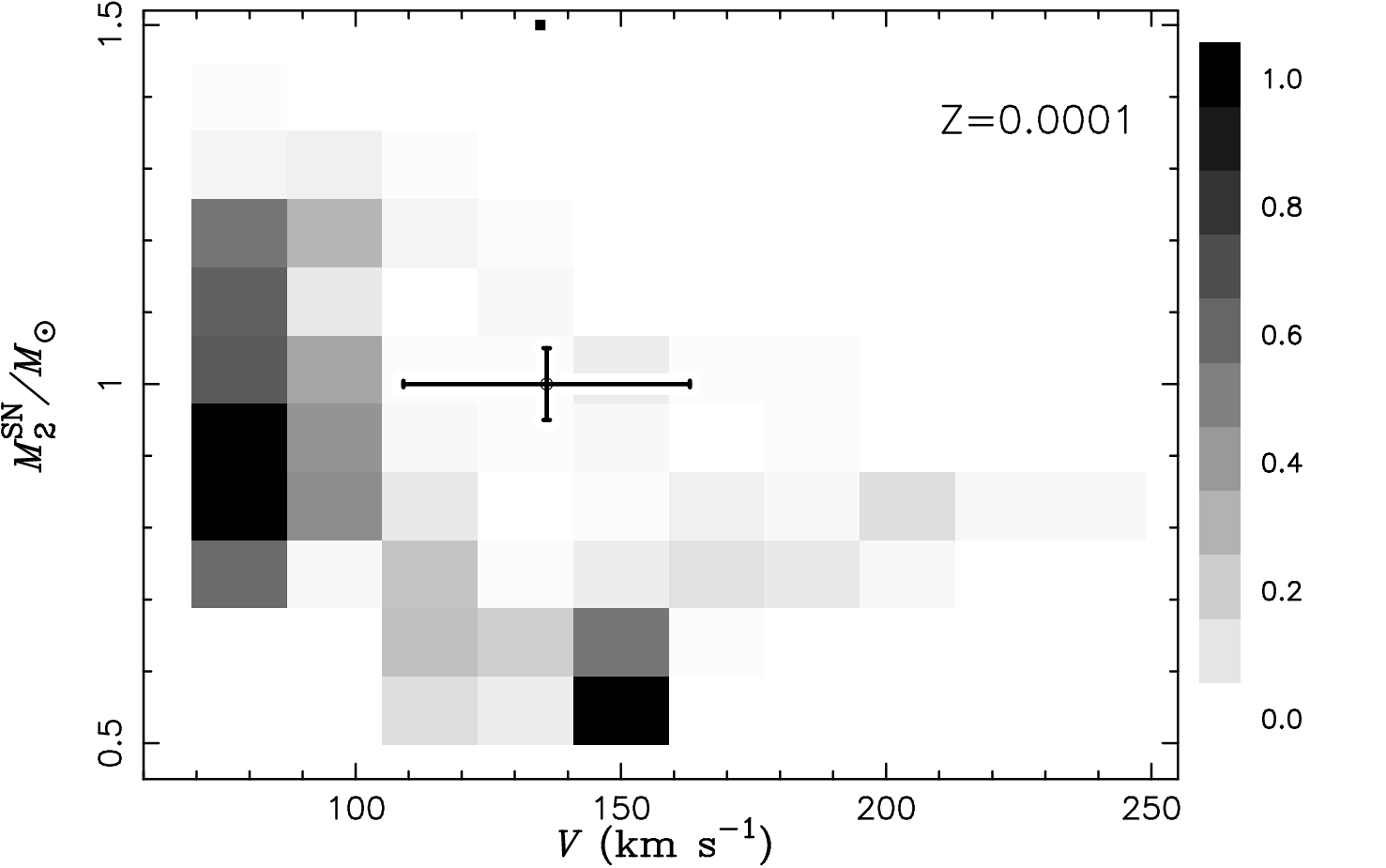,width=0.3\textheight,height=0.40\textwidth,bbllx=60pt,bblly=100pt,bburx=585pt,bbury=676pt,angle=270}}
 \suppressfloats[t]
 \caption{Similar to Fig \ref{vorbmms0302} but for $Z=0.0001$. Top: $\alpha_{\rm
CE}=1.0$; Bottom: $\alpha_{\rm CE}=3.0$.}
 \label{vorbmms00010001}
\end{figure}

\subsection{the masses and space velocities of companions in SNe Ia remnant}\label{subs:3.3}
As mentioned in section \ref{sect:1}, in the single-degenerate
model, supernova ejecta collides into the envelope of its
companion after SN Ia explosion and strips some hydrogen-rich
material from the surface of the companion (\citealt{CHE74};
\citealt{WHE75}; \citealt{FRY81}; \citealt{TAA84};
\citealt{CHU86}; \citealt{LIV92}; \citealt{MAR00}). \citet{MAR00}
ran several high-resolution two-dimensional numerical simulation
of the collision between supernova ejecta and its companion, where
the companion is a MS star, a subgiant (SG) star, or a red giant
(RG) star. They found that the hydrogen-rich material of at least
$0.15 M_{\rm \odot}$ is stripped from the envelope of a companion.
However, \citet{MENGCH07} used a simple analytic method while a
more physical companion model than that of \citet{MAR00} to
simulate the collision and found that the stripped material may be
as low as 0.035 $M_{\odot}$. Observationally, \citet{MAT05} and
\citet{LEO07} showed that the amount of stripped hydrogen-rich
material is less than 0.02 $M_{\rm \odot}$ which can be neglected
in comparison with companion mass. After the collision, the
companion gains a kick velocity, which is much lower than the
orbital velocity of the companion (\citealt{MAR00};
\citealt{MENGCH07}).

For the reasons above, we assume that the mass of a companion is
not changed by the collision of supernova ejecta and its space
velocity after the collision is equal to its orbital velocity at
the moment of explosion. Figs. \ref{vorbmms0302} to
\ref{vorbmms00010001} present the distributions of the masses and
the space velocities of companions in SNe Ia remnant for different
metallicities. Observationally, mass can be deduced from the
spectral type of a star combining with the star's surface gravity
and luminosity, and space velocity can be obtained from proper
motion and radial velocity combining with distance. Then, our
results can be compared with observations. The cross in the
figures represents the position of Tycho G, which is a potential
candidate of the companion of Tycho's supernova (\citealt{RUI04};
\citealt{BRA04}; \citealt{HERNANDEZ09}). It is interesting that
the position of Tycho G is well consistent with our
high-metallicity results, although Tycho G was suspected not to be
the companion of Tycho's supernova (\citealt{FUH05};
\citealt{IHA07}).

We see in the figures that the masses and the space velocity both
increase with metallicity on average, which is derived from the
dependence of the final secondary mass and the final period on
metallicity as shown Figs. \ref{permms0302} to
\ref{permms00010001},

The bottom panel in Fig. \ref{vorbmms00010001} shows the ($M_{\rm
2}^{\rm SN}, V$) plane for $Z=0.0001$ and $\alpha_{\rm CE}=3.0$.
There are also two islands in the figure. The low-mass
high-velocity one is from EAGB channel and the high-mass
low-velocity island is from TPAGB channel. The reason is same to
that interpreted in subsections \ref{subs:3.1}

In these figures, we also show the position of the MS star of the recurrent nova U Sco, where its space velocity
is its orbital velocity relative to the mass center of the system.
The fate of U Sco as a SN Ia is also clearly shown in the bottom
panel of Fig. \ref{vorbmms0302}.

\begin{figure}
\centerline{\psfig{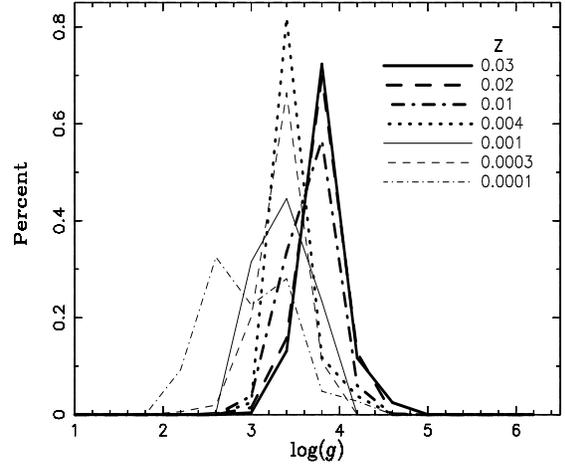}}
\caption{The distribution of the gravities of companions for
different metallicities with $\alpha_{\rm CE}=3.0$, where the
gravities are in $\rm cm/s^{\rm 2}$.} \label{fig25}
\end{figure}

\subsection{The distribution of the gravities of companions}\label{subs:3.4}
The surface gravity of the companion in a SN Ia remnant is another
parameter which can be directly obtained from spectral
observations (\citealt{RUI04}). Surface gravity will be changed by
the impact of supernova ejecta. As depicted in subsection
\ref{subs:3.2}, companions after the impact of supernova ejecta
are processing into thermal equilibrium. Meanwhile, the thermal
equilibrium of companions is still not reestablished at the moment
of supernova explosion for pre-explosion mass transfer.
Considering the insignificant change of companion mass, the
surface gravity of a companion at the moment of explosion may
represent a real one after the interaction to some extent, and
then we could make a comparison between the distribution of the
calculated surface gravities of companions and observations. We
show the distributions of the surface gravities at the moment of
explosion for various $Z$ and $\alpha_{\rm CE}=3.0$ in Fig.
\ref{fig25}. $\alpha_{\rm CE}$ does not significantly change the
distributions expect for $Z=0.0001$. Double peaks also appear for
the case of $Z=0.0001$ in the figure. The low-gravity peak is from
the TPAGB channel and the high-gravity one is from the EAGB
channel. We see from the figure that the surface gravity for
high-metallicity stars are generally larger than that for
low-metallicity ones, which is directly derived from the larger
masses and smaller radii of the companions with a high $Z$. Most
stars have surface gravity $\log g$ in the range of 2-5. The
potential companion of Tycho's supernova, Tycho G, has a surface
gravity $\log g$ between $3.0$ and $4.0$ (\citealt{RUI04}). Seen
from the figure, it is difficult to judge the metallicity of Tycho
G based on its surface gravity.

\begin{figure}
\centerline{\psfig{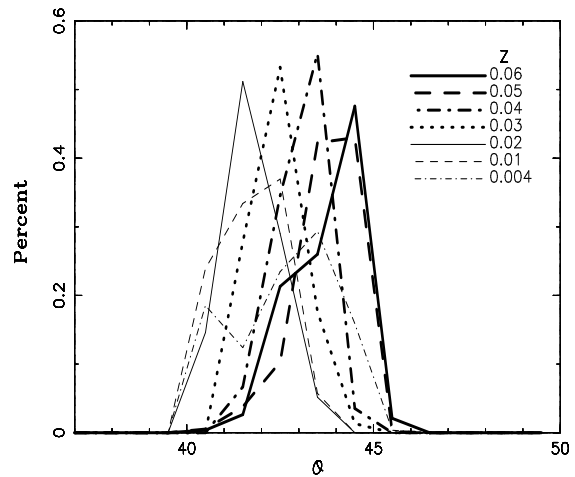}}
\caption{The distribution of the opening angles of the holes in
SNe Ia remnant for different metallicities with $\alpha_{\rm
CE}=3.0$.} \label{fig26}
\end{figure}
\subsection{Percentage of SNe Ia with polarization spectrum}\label{subs:3.5}
After the interaction between supernova ejecta and its companion,
the companion star carves out a conical hole of opening angle
$30^{\circ}$ - $40^{\circ}$ in the supernova ejecta and the hole
will never disappear because the ejecta are moving supersonically
(\citealt{MAR00}). The aspheric configuration of supernova ejecta
may reveal itself by polarization spectrum near maximum light
(\citealt{KAS04}). With the advance of spectropolarimetric
observations, the nature of SN Ia asphericity becomes an important
relevant test of the single degenerate progenitor scenario
(\citealt{WANGLF08}). \citet{KAS04} studied the effect of the hole
on the spectra of SN Ia near maximum light and found that if the
opening angle is larger than $20^{\circ}$, polarization spectrum
may be obtained from any viewing angle. Considering the limit of
spectropolarimetric observations (polarization should be larger
than 0.2\%, \citealt{LEO05}), a polarization spectrum can be
detected when viewing angle is smaller than $120^{\circ}$
(defining that the viewing angle down the hole is $0^{\circ}$).

The opening angle of the hole, $\theta$, in a SN Ia remnant is
mainly determined by $A/R_{\rm 2}^{\rm SN}$ (\citealt{MAR00}). We
use
\begin{equation}
\theta=66.82-9.285(A/R_{\rm 2}^{\rm SN})+0.3784(A/R_{\rm 2}^{\rm
SN})^{\rm 2}
\end{equation}
to calculate the opening angle of the hole. The equation is fitted
from the half-mass angle of the distribution of stripped material
given by \citet{MAR00} and the error of the equation is less than
1\%. Note that the half-mass angle may be different from the
opening angle and equation (6) may overestimate the opening angle
by about $5^{\circ}$. Here, we assume that equation (6) is valid
for all metallicities since metallicity may not significantly
affect the interaction between supernova ejecta and its companion
(\citealt{MENGCH07}).

Fig. \ref{fig26} show the distribution of the opening angles of
the holes in SNe Ia remnants for different metallicities with
$\alpha_{\rm CE}=3.0$. The results of $\alpha_{\rm CE}=1.0$ are
similar to those of $\alpha_{\rm CE}=3.0$ expect for the case of
$Z=0.0001$. Double peaks also appear in Fig. \ref{fig26}, where
the high-$\theta$ peak is from the TPAGB channel and the
low-$\theta$ peak is from the EAGB channel. We see in the figure
that the opening angles of the holes in all SNe Ia remnants are
always larger than $20^{\circ}$ even if the overestimate by
equation (6) is considered. We assume that only when the view
angle is smaller than $120^{\circ}$, polarization can be detected.
Then, the percentage of SNe Ia with polarization spectrum is
$1-\frac{1-\cos 60^{\circ}}{2}=0.75$.

\section{DISCUSSIONS}\label{sect:4}
\subsection{Tycho G}\label{subs:4.1}
It is believed that SN Ia is from the thermonuclear runaway of a
CO WD in a binary system. The CO WD accretes material from its
companion to increase its mass. When its mass reaches its maximum
stable mass, it explodes as a thermonuclear runaway and almost
half of the WD mass is converted into radioactive nickel-56
(\citealt{BRA04}). Two progenitor models of SNe Ia have competed
for about three decades. One is a single-degenerate model, which
is widely accepted (\citealt{WI73}). In this model, a CO WD
increases its mass by accreting hydorgen- or helium-rich matter
from its companion, and explodes when its mass approaches the
Chandrasekhar mass limit. The companion may be a main-sequence
star (WD+MS) or a red-giant star (WD+RG) (\citealt{YUN95};
\citealt{LI97}; \citealt{HAC99a, HAC99b}; \citealt{NOM99};
\citealt{LAN00}). Between the two channels, WD + MS model is
widely studied and some observations also uphold the channel
(\citealt{HAC99a, HAC99b}; \citealt{NOM99}; \citealt{LAN00};
\citealt{HAN04}; Paper I). \citet{HK03a, HK03b} suggested that
supersoft X-ray sources (SSSs) may be good candidates for the
progenitors of SNe Ia, where some of SSSs belong to WD+MS channel.
Recently, \citet{HK05, HK06a, HK06b} and \citet{HKL07} showed that
several recurrent novae is possibly the progenitor of SNe Ia and
some of them belong to the WD+RG channel. Observationally,
\citet{PAT07} suggested that the companion of the progenitor of SN
2006X were an early RGB star. However, \citet{HKN08} argued a WD +
MS nature for this SN Ia. Considering a smaller calculated
Galactic birth rate of SNe Ia from WD + MS channel than that
derived observationally (\citealt{HAN04}; Paper I) and some WD +
RG systems as the candidates of SNe Ia progenitors
(\citealt{HK06a,HK06b}; \citealt{HKL07}; \citealt{PAR07}), WD + RG
channel should be carefully investigated further although some BPS
results showed a small contribution of WD + RG channel to the
birth rate of SNe Ia (\citealt{YUN98}; \citealt{HAN04}). The other
progenitor model of the SNe Ia is a double degenerate model (DD,
\citealt{IT84}; \citealt{WEB84}), in which a system consisting of
two CO WDs loses orbital angular momentum by gravitational wave
radiation and merges. The merger may explode if the total mass of
the system exceeds the Chandrasekhar mass limit (see the reviews
by \citealt{HN00} and \citealt{LEI00}). The birth rate calculated
from this channel was comparable with the observational rate
(\citealt{HAN98}; \citealt{YUN98, YUN00}; \citealt{TUT02}). SN
2003fg and SN 2005hj is likely the cases from the DD channel
(\citealt{HOW06}; \citealt{BRA06}; \citealt{QUI07}). In addition,
KPD 1930+2752 may be an excellent candidate of DD SN Ia
progenitor, whose total mass ($\sim1.52 M_{\rm \odot}$) exceeds
the Chandrasekhar mass limit and whose orbital shrinkage caused by
gravitational wave radiation will lead to the merger of the binary
in about 200 Myr, much smaller than the Hubble time
(\citealt{GEI07}). However, please pay attention that
\citet{ERGMA01} argued that, from detailed binary evolution
calculation, the final mass of KPD 1930+2752 is smaller than the
Chandrasekhar mass limit due to a large amount of mass loss during
evolution.

A good way of discriminating between the many SN Ia progenitor
scenarios is to search the companion of a SN Ia in its remnant.
Unless the companion is another WD (DD channel, in which it has
been destroyed by the mass-transfer process itself before
explosion), it survives and shows some special properties in its
spectra, which is originated from the contamination of supernova
ejecta (\citealt{MAR00}; \citealt{RUI04}; \citealt{BRA04}).
Tycho's supernova, which is one of only two SNe Ia observed in our
Galaxy, provides an opportunity to address observationally the
identification of the surviving companion. \citet{RUI04} searched
the region of the remnant of Tycho's supernova and suggested that
Tycho G, a sun-like star, is the companion of Tycho's supernova.
Although, \citet{IHA07} argued that the spectrum of Tycho G does
not show any special properties, which seems to exclude the
possibility of Tycho G to be the companion of Tycho' supernova,
the analysis of the chemical abundances of the Tycho G upholds the
companion nature of the Tycho G (\citealt{HERNANDEZ09}).

Interestingly, some integral properties of Tycho G, i.e. the mass,
the space velocity, the radius and the surface gravity, are all
consistent with our binary population synthesis results (see Figs.
\ref{ratrsn0302}, \ref{vorbmms0302} and \ref{fig25}). Then, Tycho
G is very likely to be the companion of Tycho's supernova
(see also \citealt{MENGYG09}). If Tycho G were the companion of
Tycho's supernova as shown by our BPS results, it would challenge
one's understanding about the physics of SNe Ia, such as the
interaction between SNe Ia ejecta and companions.

\subsection{The simulation of the interaction between SNe Ia ejecta and companions}\label{subs:4.2}
In section \ref{sect:1}, we have shown that \citet{MAR00} ran
several high-resolution two-dimensional numerical simulations of
the collision between the ejecta and the companion. They claimed
that about $0.15-0.17 M_{\odot}$ of hydrogen-rich material is
stripped from a MS or a SG companion. They also found that
in a sense of the collision, there is no difference whatever the
companion is a MS star or a SG star. However, their results did
not obtain confirmation by observations (\citealt{MAT05};
\citealt{LEO07}). \citet{MENGCH07} used a simple analytic method
to simulate the the interaction between SNe Ia ejecta and
companions and found that for a given condition, more
hydrogen-rich material is stripped from the envelope of a SG
companion than that of a MS companion. To discuss the validity of
their method, they repeated their work using the same method in
their paper and the companion model in \citet{MAR00}, and found
that the result of SG model is similar to that in \citet{MAR00},
while the amount of the stripped hydrogen-rich material of MS
model is much smaller than that in \citet{MAR00}. The reason of
the difference is that the SG model in \citet{MAR00} is from
\citet{LI97}, who used a method and a stellar evolution
code similar to that in Paper I. However, the MS model in
\citet{MAR00} is a solar model, which is not a typical case (see
Figs \ref{ratrsn0302} to \ref{ratrsn00010001}). \citet{MENGCH07}
claimed that the difference of the results between their analytic
method and the numerical simulation in \citet{MAR00} is mainly
derived from the different stellar structure of the companion.
Thermal equilibrium is not reestablished for the companion star at
the moment of supernova explosion since the mass transfer is still
processing before supernova explosion. Additionally, \citet{MAR00}
and \citet{MENGCH07} argued that the luminosity of a companion
after the impact of SNe Ia ejecta would rise sharply to about 5000
$L_{\odot}$, which is too high to compare with that of Tycho G.
Therefore, considering that the properties of Tycho G is
consistent with our BPS results, a detailed numerical simulation
about the interaction between supernova ejecta and its companion
should be encouraged by a more physical companion model than that
in \citet{MAR00}.

Recently, \citet{PAK08} used a more physical companion model and
similar numerical simulation to that in \citet{MAR00} to
recalculate the interaction between supernova ejecta and
companion. They found a similar results to that in \citet{MAR00},
and at the same time they claimed that under some special
situation, results consistent with observation may be obtained,
and then they claimed that theory does not conflict with
observation. Based on the result in \citet{PAK08}, if $A/R_{\rm
2}^{\rm SN}$ is smaller than 6, the amount of striped mass from
companions should be larger than 0.02 $M_{\odot}$. According to
the results in this paper (see figures 5 to 8), $A/R_{\rm 2}^{\rm
SN}$ is always smaller than 6, and then the range of the stripped
mass is from 0.07 $M_{\odot}$ to 0.16 $M_{\odot}$, which is
consistent with the results of \citet{MAR00} and \citet{MENGCH07}.
The effort of \citet{PAK08} then do not overcome the confliction
between theory and observations (\citealt{MAT05};
\citealt{LEO07}). The impact of a SN Ia on its companions still
should be studied carefully. Maybe, unsymmetrical explosion plays
an important role (\citealt{PLE04}).

\subsection{Polarization of SNe Ia}\label{subs:4.3}
As an important diagnostic tool for discriminating among SN Ia
progenitor systems and theories of the explosion physics,
spectropolarimetry provides the direct probe of early-time SN
geometry. The essential idea is as follows: electron scattering
dominates a hot young SN atmosphere and its nature is highly
polarizing. For an unresolved source with a spherical distribution
of scattering electrons, the directional components of the
electric vectors of the scattered photons cancel exactly, yielding
zero net linear polarization. An incomplete cancellation will be
derived from any asymmetry in the distribution of the scattering
electrons, or of absorbing material overlying the
electron-scattering atmosphere. Then, a net polarization is
resulted (\citealt{LF05}; \citealt{WANGLF08}). Single-degenerate
model provides a natural way to produce the asymmetry. The exist of
a companion may change the configuration of supernova ejecta and a
polarization spectrum is expected. In this paper, we use the
results of \citet{MAR00} and \citet{KAS04} to calculate the
percentage of SNe Ia with polarization spectrum and found that
about 75\% of all SN Ia may be detected by spectropolarimetry.
However, this result critically depends on the following
assumptions: (i) all SNe Ia are from single-degenerate progenitor
systems, (ii) \citet{MAR00} showed reliable simulations in the
sense that a hole is indeed formed and does not quickly close with
time and (iii) \citet{KAS04} provided a reasonable simulation of
polarization spectrum resulted from the exist of a hole in a
supernova ejecta.

It is likely that the single-degenerate model is only one of the
reliable models, such as the prompt component in the two-component
model suggested by \citet{SB05} and \citet{MAN06}. At present, any
definitive conclusion about DD model is premature, and this
scenario can naturally result in an asymmetry of distribution of
supernova ejecta. One mechanism is the rapid rotation of a WD
before supernova explosion, which leads to the change of the
stellar shape. Another one is that there may be a thick accretion
disk around CO WD and the disk is an origin of the asymmetry of
the configuration of supernova ejecta (see the reviews by
\citealt{HN00} and \citealt{LEI00} for details about DD model).
Additionally, explosion mechanism itself may also produce the
asymmetry and a polarization spectrum is expected
(\citealt{PLE04}; \citealt{KP05}). Then, it is probable that the
percentage of SNe Ia with polarization spectrum estimated in this
paper is a lower limit. At present, almost all SNe Ia, which are
observed by spectropolarimetry, had various degrees of
polarization signal (\citealt{LEO05}).

\subsection{The percentage of SN 1991T-like}\label{subs:4.3}
SN 1991T is an overluminous event and has a rather broad light
curve ($\Delta m_{\rm 15}=0.95\pm0.05$, where $\Delta m_{\rm 15}$
is the magnitude difference between its maximum and 15 days later
\citealt{PHI99}), which is often taken as an indication of a large
$^{\rm 56}$Ni mass (\citealt{HOF95}; \citealt{NUG97};
\citealt{PE01}). Its spectrum also showed some special properties,
i.e. dominated by FeII and FeIII lines at maximum light, while the
spectrum of a normal SN Ia is dominated by SiII line. A large
$^{\rm 56}$Ni mass may well explain the peculiar spectral
appearance (\citealt{JEF92}; \citealt{MAZZA95}). Nevertheless,
\citet{KAS04} suggested a second, physically very different route
to explain the spectral peculiarities of SN 1991T
--- one could be peering down an ejecta hole. They found that their
synthesis spectrum down the ejecta hole can be comparable with
that of SN 1991T. We calculate the frequency of the special event
for different metallicities by assuming that if view angle is
smaller than the opening angle of the hole in a SN Ia ejecta, the
SN Ia show properties of SN 1991T-like. The opening angel is from
Fig \ref{fig26}. The calculated birth rate of SN 1991T-like
supernovae is from 13\% to 14\% for different metallicities. The
rate slightly increase with $Z$ and is a little larger than the
estimation by \citet{KAS04} ($\sim12\%$). Both \citet{BRA01} and
\citet{LI01} gave the observed rate of SN 1991T-like from 3\% to
5\%. If taking SN 1999aa-like as SN 1991T-like events, the birth
rate of SN 1991T/SN 1999aa-like is $20\%\pm7\%$ (\citealt{LI01}).
Our results match with that of \citet{LI01} within errors. So, it
is possible that SNe 1991T-like have not any special properties in
physics \textbf{except for} the viewing angle of an observer.

\section{SUMMARY AND CONCLUSION}\label{sect:5}
Incorporating the results of \citet{MENGCH09} into Hurley's rapid
binary evolution code, we have carried out a series of binary
population synthesis calculation and systemically study the
properties of the companions of SNe Ia for different $Z$ at the
moment of explosions. We give the distributions of the masses,
$M_{\rm 2}^{\rm SN}$, the radii, $R_{\rm 2}^{\rm SN}$, of
companions and the periods, $P_{\rm SN}$ and the ratios of
separations to radii, $A/R_{\rm 2}^{\rm SN}$, of final binary
systems for various $Z$ at the moment of supernova explosion and
find that generally, $M_{\rm 2}^{\rm SN}$ increases and $R_{\rm
2}^{\rm SN}$ and$P_{\rm SN}$ decrease with $Z$, while the
distributions of $A/R_{\rm 2}^{\rm SN}$ are similar for all
metallicities. These parameters can be applied to constrain the
numerical simulation of the interaction between the ejecta of a
supernova and its companion. In addition, these parameters can
help to judge whether a WD + MS system may explode as a SN Ia or
not. We also show the distributions of some integral properties,
i.e. the masses, the radii, the surface gravities and the space
velocities of companions for different $Z$ after the interaction.
The distributions can provide help to search companion in
supernova remnant. Especially, some integral properties of Tycho G
(a potential candidate of the companion of Tycho's supernova),
such as mass, radius, surface gravity and space velocity, well
match with our BPS results. This fact may challenge our
understanding about the physics of SNe Ia, especially the
interaction between supernova ejecta and its companion. Using the
results simulated by \citet{MAR00} and \citet{KAS04}, we find that
about 75\% of all supernovae can be detected by
spectropolarimetric observations. If considering that there may be
different progenitor models of SNe Ia and different explosion
mechanisms, the percentage could increase.


\label{lastpage}

\end{document}